\documentclass[12pt]{article}
\title{\normalsize{The causal meaning of Fisher's average effect}}
\date{}
\author
{\normalsize{James J. Lee}$^{1\ast}$\\
\normalsize{Carson C. Chow}$^{1}$\\
\\
\normalsize{$^{1}$Laboratory of Biological Modeling}\\
\normalsize{National Institute of Diabetes and Digestive and Kidney Diseases}\\
\normalsize{National Institutes of Health}\\
\normalsize{Bethesda, MD 20892, USA}\\
\\
\normalsize{$^\ast$To whom correspondence should be addressed; E-mail:  leejj5@mail.nih.gov.}
}

\usepackage[american]{babel}
\usepackage{array}
\usepackage{tikz}
\usetikzlibrary{matrix}
\usepackage{amsmath,amssymb,bm}
\usepackage{csquotes}
\usepackage[style=authoryear-comp,sortcites=false,sorting=nyt,firstinits=true,maxcitenames=2,natbib=true,maxbibnames=99,dashed=false]{biblatex}
\DeclareLanguageMapping{american}{american-apa}
\DeclareNameAlias{sortname}{last-first}

\renewbibmacro*{name:andothers}{
  \ifboolexpr{
    test {\ifnumequal{\value{listcount}}{\value{liststop}}}
    and
    test \ifmorenames
  }
    {\ifnumgreater{\value{liststop}}{1}
       {\finalandcomma}
       {}%
     \andothersdelim\bibstring[\emph]{andothers}}
    {}}

\DeclareFieldFormat[article]{title}{#1\isdot}
\DeclareFieldFormat[inproceedings]{title}{#1\isdot}
\DeclareFieldFormat[unpublished]{title}{#1\isdot}
\DeclareFieldFormat*{volume}{\mkbibbold{#1}}
\DeclareFieldFormat{pages}{#1}
\renewbibmacro{in:}{}

\AtEveryBibitem{\clearfield{doi}}
\AtEveryBibitem{\clearfield{url}}
\AtEveryBibitem{\clearfield{number}}
\AtEveryBibitem{\clearfield{month}}
\AtEveryBibitem{\clearfield{chapter}}

\def\citeapos#1{\citeauthor{#1}'s (\citeyear{#1})}

\newlength{\arrowsize}  
\pgfarrowsdeclare{biggertip}{biggertip}{  
  \setlength{\arrowsize}{0.4pt}  
  \addtolength{\arrowsize}{.5\pgflinewidth}  
  \pgfarrowsrightextend{0}  
  \pgfarrowsleftextend{-5\arrowsize}  
}{  
  \setlength{\arrowsize}{0.4pt}  
  \addtolength{\arrowsize}{.5\pgflinewidth}  
  \pgfpathmoveto{\pgfpoint{-5\arrowsize}{4\arrowsize}}  
  \pgfpathlineto{\pgfpointorigin}  
  \pgfpathlineto{\pgfpoint{-5\arrowsize}{-4\arrowsize}}  
  \pgfusepathqstroke  
}  

\begin{document}
\hyphenation{dis-equil-i-brium pheno-types pheno-type geno-type
geno-types counter-intuitive inter-action}

\maketitle

RESEARCH PAPER \\

RUNNING HEAD: Causal meaning of average effect

\baselineskip24pt

\newpage

\begin{center}
Summary
\end{center}

In order to formulate the Fundamental Theorem of Natural Selection, Fisher defined the \emph{average excess} and \emph{average effect} of a gene substitution. Finding these notions to be somewhat opaque, some authors have recommended reformulating Fisher's ideas in terms of covariance and regression, which are classical concepts of statistics. We argue that Fisher intended his two averages to express a distinction between correlation and causation. On this view the average effect is a specific weighted average of the actual phenotypic changes that result from physically changing the allelic states of homologous genes. We show that the statistical and causal conceptions of the average effect, perceived as inconsistent by Falconer, can be reconciled if certain relationships between the genotype frequencies and non-additive residuals are conserved. There are certain theory-internal considerations favoring Fisher's original formulation in terms of causality; for example, the frequency-weighted mean of the average effects equaling zero at each locus becomes a derivable consequence rather than an arbitrary constraint. More broadly, Fisher's distinction between correlation and causation is of critical importance to gene-trait mapping studies and the foundations of evolutionary biology. \\

Keywords: quantitative genetics, causality, confounding, selection bias, natural selection

\newpage

\section*{1. Introduction}

Darwin perceived that hereditary variation in fitness leads to an increase in adaptive complexity. In an attempt to provide a Mendelian and mathematical formulation of this profound insight, Fisher expounded the Fundamental Theorem of Natural Selection (FTNS), which in a modern paraphrase states that the partial increase in population mean fitness ascribable solely to changes in allele frequencies by natural selection is equal to the additive genetic variance in fitness \citep{bennett:1956,kimura:1958,price:1972,ewens:1989,ewens:2004,ewens:2011,frank:1992,edwards:1994,edwards:2002,lessard:1997,okasha:2008}. In the discrete-time formulation of the FTNS, the additive genetic variance is proportional to this partial increase, as it must be divided by the mean fitness.

In his exposition of the FTNS, Fisher took some pains to define the concepts of
\emph{average excess} and \emph{average effect}. In his own words,
\begin{quote}
Let us now consider the manner in which any quantitative individual measurement, such as human stature, may depend upon the individual genetic constitution. We may imagine, in respect of any pair of alternative [alleles], the population divided into two portions, each comprising one homozygous type together with half of the heterozygotes, which must be divided equally between the two portions. The difference in average stature between these two groups may then be termed the \emph{average excess} (in stature) associated with the gene substitution in question\ldots. \citep[p.~30, emphasis added]{fisher:1930:gtns}
\end{quote}
In contrast,
\begin{quote}
[b]y whatever rules \ldots the frequency of different gene combinations, may be governed, the substitution of a small proportion of the genes of one [allelic] kind by the genes of another will produce a definite proportional effect upon the average stature. The amount of the difference produced, on the average, in the total stature of the population, for each such gene substitution, may be termed the \emph{average effect} of such substitution, in contra-distinction to the average excess as defined above\ldots. \citep[p.~31, emphasis added]{fisher:1930:gtns}

It is natural to conceive [of the average effect] as the actual increase in the total of the measurements of a population, when without change in the environment, or the mating system, the gene substitution is \emph{experimentally} brought about, as it might be by mutation. \citep[p.~373, emphasis added]{fisher:1941}
\end{quote}

This paper addresses a puzzle raised by \citet{falconer:1985} in his brilliant explication of Fisher's two genetic averages. Falconer assumed that what Fisher meant by the quoted definition of the average effect was as follows. We randomly sample a zygote immediately after fertilization but before the onset of any developmental events. If the zygote's genotype contains a gene of a certain allelic type, say $\mathcal{A}_1$, we change it to $\mathcal{A}_2$. This experimental intervention may lead to a value of the focal phenotype at the time of measurement that differs from what it would have been if the intervention had not been performed. Falconer reasoned that the expected magnitude of this difference corresponds to Fisher's verbal definition of the average effect.

Falconer then showed that \citeapos{fisher:1941} now widely accepted mathematical definition of the average effect---the partial regression coefficient of gene count in the linear regression of the phenotype on all loci in the genome---does not generally coincide with the definition in terms of experimental gene substitutions performed at random. Falconer expressed surprise at the apparent invalidity of the latter definition, given that ``Fisher uses the imaginary replacement of one allele by another as a verbal description to introduce the idea of average effect, and it seems to have been seen by him as the basis for the concept'' (p.~334).

Falconer correctly perceived the importance of experimental intervention to Fisher's conception of the average effect. Indeed, Fisher did not even bother to spell out his regression definition in the first edition of \emph{The Genetical Theory of Natural Selection}. Furthermore, to any reader familiar with Fisher's work on experimental design and his controversial stance on the tobacco-cancer connection, the quotations given above must bring into mind his repeated admonition that an observed \emph{excess} in the average measurement of one group over another can always be interpreted as the causal \emph{effect} of the factor distinguishing the groups under the following circumstance: the allotment of members to groups has been randomized in a controlled experiment \citep{fisher:1935,fisher:1958:cancer}. This preoccupation with causation is one of the stark contrasts between Fisher and his nemesis Karl Pearson; contrary to the intellectual fashion of the Edwardian era, Fisher did not regard causality as a meaningless concept. In the inaugural issue of the journal \emph{Philosophy of Science}, the word \emph{cause} and its derivatives appear in \citet{fisher:1934} no fewer than seventy times. Over much resistance by seasoned experimenters \citep{box:1978}, Fisher advocated randomization in experimental design for the precise purpose of distinguishing causation from spurious correlations brought about by confounding variables. There is thus compelling reason to believe that the notion of experimental control revealing causation is critical to the proper interpretation of the average effect.

We argue that a more nuanced reading of Fisher's writings can bring his experimental and regression definitions of the average effect into full agreement in certain special cases. We then provide reasons to favor the experimental definition in more general situations. A striking disadvantage of the regression definition is that its use invalidates the FTNS if some of the variance in fitness has environmental causes. 

For simplicity our main text mostly follows \citet{falconer:1985} in treating the case of a single locus with two alleles. We provide the generalization to multiple alleles and loci in two of the later sections. Some interesting new concepts do arise in this generalization, but the central ideas can be conveyed without multilocus notation, which seems inevitably to be either cumbersome or opaque.

\section*{2. A Notation for Causal Notions}

A formal symbolic language to distinguish causal relations from merely correlational ones, such as the counterfactual notation of \citet{neyman:1923} and \citet{rubin:2005}, was not to our knowledge ever adopted by Fisher. This is despite the fact that he frequently wrote about this distinction. Although such formalisms lack the elegance of Fisher's prose, adopting the appropriate formalism is an aid to understanding.

For this purpose we adopt the $do$ operator of \citet{pearl:1995}. We are to interpret an expression such as $\mathbb{E}[Y \,|\, do(x)]$ to mean the expectation of $Y$ given that the random variable $X$ has been \emph{experimentally fixed} to the value $x$. The contrast between conditional quantities containing the $do$ symbol and traditional conditional quantities is evident in the expressions
\begin{equation}\label{rainAmud}
\mathbb{P}(mud \,|\, rain) \ge \mathbb{P}(mud) \quad \textrm{and} \quad \mathbb{P}(rain \,|\, mud) \ge \mathbb{P}(rain)
\end{equation}
and
\begin{equation}\label{rainCmud}
\mathbb{P}[mud \,|\, do(rain)] \ge \mathbb{P}(mud) \quad \textrm{and} \quad \mathbb{P}[rain \,|\, do(mud)] = \mathbb{P}(rain).
\end{equation}
(\ref{rainAmud}) indicates that we are more likely to find mud if we have already observed rain. Because co-occurrence is symmetric, it also becomes more likely that it has rained if we have already observed mud. On the other hand, (\ref{rainCmud}) symbolizes the much stronger and asymmetrical assertion that rain causes mud and not \emph{vice versa}; muddying up the backyard with a garden hose will not make it rain.

This notation and its associated machinery may be of some benefit in the burgeoning field of genome-wide association studies (GWAS), where it is important to single out genetic variants with a causal effect on a given phenotype from markers that are merely associated with the phenotype for other reasons, including linkage disequilibrium (LD) with a nearby causal variant \citep{visscher:2012}. Letting $Y$ denote the phenotype of interest, we can say that a genetic variant is a causal variant if the equality
\begin{equation}\label{defcause}
\mathbb{E}[Y \,|\, do(\mathcal{A}_1 \mathcal{A}_1)] = \mathbb{E}[Y \,|\, do(\mathcal{A}_1 \mathcal{A}_2)] = \mathbb{E}[Y \,|\, do(\mathcal{A}_2 \mathcal{A}_2)] 
\end{equation}
does not hold. The expectation is taken over the space of all possible multilocus genotypes and environments. Note that the equality does in fact hold for a non-causal marker locus in LD with a causal locus. If we could experimentally mutate a randomly chosen zygote's genotype at a biologically inert marker locus immediately before the onset of development, we would not expect any ensuing change in the phenotype.

The $do$ notation is more than a convenient means of fixing ideas. The treatise of \citet{pearl:2009} grounds this symbol in a rich syntax and semantics. From one point of view, the work of Pearl can be regarded as a vast generalization of \citeapos{wright:1968} path analysis.

For simplicity we will speak of events in the life cycle such as fertilization, development, and phenotypic measurement as if all individuals experienced each such event at the same time---a convention that is appropriate for an organism with a life cycle consisting of discrete and non-overlapping generations. We can then speak of selecting one zygote for an experimental treatment from all those zygotes making up the current generation. Our discussion also applies, however, to organisms with a life cycle consisting of continuous and overlapping generations. In this case a quantity such as $\mathbb{E}[Y \,|\, do(\mathcal{A}_1 \mathcal{A}_1) ]$ is to be interpreted as the present phenotypic value that a randomly selected organism would have been expected to obtain if its genotype could have been converted to $\mathcal{A}_1 \mathcal{A}_1$ immediately after its own fertilization. Fisher's own writings suggest the importance of counterfactual thinking. In a summary of his work on the correlations between relatives, he wrote: ``[I]t should be clearly understood what we mean by a \emph{cause of variability}. If we say, `This boy has grown tall because he has been well fed,' we are not merely tracing out cause and effect in an individual instance; we are suggesting that he might quite probably have been worse fed, and that in this case he would have been shorter'' \citep[p.~214, emphasis in original]{fisher:1919}. The $do$ operator bears both interventional and counterfactual interpretations. If necessary, each organism can be weighted by reproductive value.

\section*{3. Falconer's Interpretation of the Experimental Average Effect}

We can use the $do$ operator to symbolize the gene substitutions in Fisher's thought experiment. Here we use it to review Falconer's understanding of this experiment for a single biallelic locus. We first note that if genotypic and environmental causes of phenotypic variation act additively and independently, then quantities such as $\mathbb{E}(Y \,|\, \mathcal{A}_1 \mathcal{A}_1)$ are precisely equal to $\mathbb{E}[Y \,|\, do(\mathcal{A}_1 \mathcal{A}_1)]$ at the single causal locus. Until we say otherwise, we assume the stochastic independence of genotypes and environments.

Following the notation of \citet{fisher:1918}, we let $P$, $2Q$, and $R$ denote the respective frequencies of the genotypes $\mathcal{A}_1 \mathcal{A}_1$, $\mathcal{A}_1 \mathcal{A}_2$, and $\mathcal{A}_2 \mathcal{A}_2$. Given that a zygote's genotype is $\mathcal{A}_1 \mathcal{A}_1$, we write the expected phenotypic effect of changing a gene's allelic type from $\mathcal{A}_1$ to $\mathcal{A}_2$ as
\begin{equation}\label{change1}
\Delta Y \,|\, \mathcal{A}_1 \mathcal{A}_1 \rightarrow \mathcal{A}_1 \mathcal{A}_2 = \mathbb{E}[Y \,|\, do(\mathcal{A}_1 \mathcal{A}_2), \mathcal{A}_1 \mathcal{A}_1] - \mathbb{E}(Y \,|\, \mathcal{A}_1 \mathcal{A}_1).
\end{equation}
There is no contradiction in conditioning on both the observation of $\mathcal{A}_1 \mathcal{A}_1$ and the experimental setting of the genotype to $\mathcal{A}_1 \mathcal{A}_2$. This simply means that instead of performing the experiment on a zygote sampled at random from the entire population, we perform it specifically on a zygote that would otherwise have borne the genotype $\mathcal{A}_1 \mathcal{A}_1$. Similarly, we define
\begin{equation}\label{change2}
\Delta Y \,|\, \mathcal{A}_1 \mathcal{A}_2 \rightarrow \mathcal{A}_2 \mathcal{A}_2 = \mathbb{E}[Y \,|\, do(\mathcal{A}_2 \mathcal{A}_2), \mathcal{A}_1 \mathcal{A}_2] - \mathbb{E}(Y \,|\, \mathcal{A}_1 \mathcal{A}_2).
\end{equation}

The problem with identifying \emph{the} effect of a gene substitution---as in identifying the effect of an alteration to any nonlinear causal system---is that the expected change depends on the context. In other words (\ref{change1}) and (\ref{change2}) are not equal in general. Falconer supposed that Fisher arrived at the ``average effect'' of substituting $\mathcal{A}_2$ for $\mathcal{A}_1$ by averaging (\ref{change1}) and (\ref{change2}) in the following way. We sample a zygote at random and then select one of its genes at random. If the chosen gene is of allelic type $\mathcal{A}_2$, we leave it alone. If the chosen gene is of type $\mathcal{A}_1$, we change it to $\mathcal{A}_2$. The expected phenotypic effect of the gene substitutions performed under this scheme is thus
\begin{equation}\label{falconer}
\frac{P (\Delta Y \,|\, \mathcal{A}_1 \mathcal{A}_1 \rightarrow \mathcal{A}_1 \mathcal{A}_2) + Q (\Delta Y \,|\, \mathcal{A}_1 \mathcal{A}_2 \rightarrow \mathcal{A}_2 \mathcal{A}_2) }{P + Q}.
\end{equation}

Falconer pointed out that (\ref{falconer}) does not agree with the regression definition of the average effect that \citet{fisher:1941} gave in an article criticizing Sewall Wright for conflating the average excess and average effect. This article required explicit expressions for the two genetic averages in traditional notation, and Fisher obtained an expression for the average effect adequate for demonstrating its distinctness from the average excess by minimizing the sum of squares
\begin{equation}\label{regress}
P [ \mathbb{E}(Y \,|\, \mathcal{A}_1 \mathcal{A}_1) - \nu + \alpha ]^2 +
2Q [ \mathbb{E}(Y \,|\, \mathcal{A}_1 \mathcal{A}_2) - \nu ]^2 +
R [ \mathbb{E}(Y \,|\, \mathcal{A}_2 \mathcal{A}_2) - \nu - \alpha ]^2,
\end{equation}
where $\nu$ is the regression constant. Using a notation that generalizes to a locus with more than two alleles, we can express this sum of squares equivalently as
\begin{multline}\label{ewens_regress}
P [ \mathbb{E}(Y \,|\, \mathcal{A}_1 \mathcal{A}_1) - \mu - 2 \alpha_1]^2 +
2Q [ \mathbb{E}(Y \,|\, \mathcal{A}_1 \mathcal{A}_2) - \mu - \alpha_1 - \alpha_2]^2 \\
+ R [ \mathbb{E}(Y \,|\, \mathcal{A}_2 \mathcal{A}_2) - \mu - 2\alpha_2]^2 \quad
\textrm{where $\mu = \mathbb{E}(Y)$}.
\end{multline}
In the definition (\ref{regress}), then, the average effect $\alpha$ is the slope in the regression of the phenotype on gene count. $\alpha_1$ and $\alpha_2$ in (\ref{ewens_regress}) are the average effects of the two alleles individually---a notion to which we will return. For now we simply note that $\alpha$ will turn out to equal $\alpha_2 - \alpha_1$ in magnitude. There is some ambiguity in the literature over whether the outcome variable in the regression should be defined, as in (\ref{ewens_regress}), with the subtraction of the unconditional phenotypic mean \citep{fisher:1958:gtns,price:1972,ewens:2011}. However, this choice simply adds a constant term to the average effects of the individual alleles, and this term disappears in the biallelic average effect $\alpha = \alpha_2 - \alpha_1$. In our later discussion of individual average effects, we will give a compelling reason to favor the mean subtraction.

Perhaps frustrated by Fisher's concise style, Falconer concluded his article by approvingly quoting \citeapos{price:1972} remark that Fisher's ideas can be translated into well-understood concepts such as covariance and regression without dealing with his ``special'' notions of average excess and average effect.

In the following we show that the two definitions of the average effect can be reconciled, in the case of genotype-environment independence, for a specific weighting of the two possible substitutions. However, if such independence fails to hold, it is not possible to dispense with Fisher's ``special'' definition in terms of experimental gene substitutions.

\section*{4. Fisher's Experimental Average Effect}

Fisher conditioned the gene substitutions in his hypothetical experiment on the ``rules'' by which ``the frequency of different gene combinations may be governed.'' It is this difficult subtlety that Falconer did not take into account. In \emph{The Genetical Theory} Fisher's wording seems to imply that it is only the mating scheme that determines how different alleles combine to form whole-genome genotypes. Later he acknowledged that other factors also influence the departure of genotype frequencies from random combination of genes, explicitly mentioning ``the partial isolation of sections of the population'' \citep[p.~54]{fisher:1941}. The implication for the experimental gene substitutions is that they must be carried out in a manner that does not disturb the arrangement of alleles into genotypes called for by the population's rules of formation. 

The three genotype frequencies sum to unity, as do the frequencies of the two alleles. Thus, given the frequency of one allele, one more parameter is required to specify the genotype frequencies. There appears to be complete freedom in the choice of this parameter. For example, one possibility is Wright's inbreeding coefficient $F$ \citep{crow:1956}. As we later show, if we require the experimental average effect to coincide with the regression average effect in the case of genotype-environment independence, then we must choose the parameter to be $\lambda = Q^2 /(PR)$, the ratio of the squared (ordered) heterozygote frequency to the product of the homozygote frequencies. $\lambda$ can be written in the symmetrical form
\begin{equation*}
\frac{\mathbb{P}( \mathcal{A}_2 \,|\, \mathcal{A}_1 )}{ \mathbb{P}( \mathcal{A}_1 \,|\, \mathcal{A}_1 ) } \cdot
\frac{\mathbb{P}( \mathcal{A}_1 \,|\, \mathcal{A}_2 )}{ \mathbb{P}( \mathcal{A}_2 \,|\, \mathcal{A}_2 ) },
\end{equation*}
and it attains the constant value of unity if the population mates randomly, a fact first noted by \citet{hardy:1908}.

Let $p = Q + R$ denote the frequency of $\mathcal{A}_2$, and write the population mean of $Y$ as a function of allele frequency and the rules of combination, $\mu(p, \lambda)$. We now show that the expression $\mu(p + dp, \lambda) - \mu(p, \lambda)$ is proportional to the average effect, $\alpha$, obtained from regression equation (\ref{regress}). In other words the ratio $\lambda$ must be kept constant under this manipulation, \emph{whatever} the population's rules of formation have determined this ratio to be, in order for the experimental gene substitutions to yield what Fisher intended by the average effect. 

The population mean is given by the expression
\begin{equation}\label{exp_mean}
\mu= P \, \mathbb{E}[ Y \,|\, do(\mathcal{A}_1 \mathcal{A}_1)] + 2Q \, \mathbb{E}[ Y \,|\, do(\mathcal{A}_1 \mathcal{A}_2)] + R \, \mathbb{E}[ Y \,|\, do(\mathcal{A}_2 \mathcal{A}_2)].
\end{equation}
The average effect is then proportional to the change of $\mu$ with respect to $p$ while holding $\lambda$ constant. We can increase $p$ by carrying out either the intervention $\mathcal{A}_1 \mathcal{A}_1$ $\rightarrow$ $\mathcal{A}_1 \mathcal{A}_2$ or $\mathcal{A}_1 \mathcal{A}_2$ $\rightarrow$ $\mathcal{A}_2 \mathcal{A}_2$. As detailed in the Appendix, upon noting that the differential of $Q^2 = \lambda P R$ for constant $\lambda$ yields the differential equation
\begin{equation}\label{diffeq}
\frac{dP}{P} + \frac{dR}{R} = \frac{2 dQ}{Q},
\end{equation}
we find that Fisher's average effect is
\begin{equation}\label{alpha}
\alpha = \frac{c_1 (\Delta Y \,|\, \mathcal{A}_1 \mathcal{A}_1 \rightarrow \mathcal{A}_1 \mathcal{A}_2) + c_2 (\Delta Y \,|\, \mathcal{A}_1 \mathcal{A}_2 \rightarrow \mathcal{A}_2 \mathcal{A}_2) }{c_1 + c_2},
\end{equation}
where the weights are 
\begin{align*}
c_1 &= P(Q + R), \\
c_2 &= R(P + Q).
\end{align*}

Let us recapitulate the meaning of (\ref{alpha}). Immediately after fertilization we take a random sample of the zygotes bearing the genotype $\mathcal{A}_1 \mathcal{A}_1$. We then randomly assign some of these zygotes to the ``treatment,'' which consists of changing the allelic type of a gene from $\mathcal{A}_1$ to $\mathcal{A}_2$. The expected difference in phenotype between treatments and controls at the time of measurement is the causal effect of the gene substitution. We perform the analogous experiment to determine the causal effect of changing $\mathcal{A}_1 \mathcal{A}_2$ to $\mathcal{A}_2 \mathcal{A}_2$. The weighted average of the two causal effects---where the weights $c_1$ and $c_2$ are chosen so as to preserve $\lambda$ if the two types of gene substitutions are applied to the population in the ratio $c_1/c_2$---is the average effect of gene substitution holding constant the rules governing the frequencies of the different genotypes.

Now that the average effect has been defined in (\ref{alpha}), we can apply it to an example of a population changing in mean phenotypic value under a sequence of gene substitutions (Table~\ref{ex1}). This example may be seen as a numerical counterpart to the diagrammatic illustration by \citet{edwards:2002}. Suppose that the effect of changing an $\mathcal{A}_1 \mathcal{A}_1$ individual to $\mathcal{A}_1 \mathcal{A}_2$ is 3 phenotypic units, whereas the effect of changing $\mathcal{A}_1 \mathcal{A}_2$ to $\mathcal{A}_2 \mathcal{A}_2$ is $-2$. Suppose also that the numbers of the genotypes $\mathcal{A}_1 \mathcal{A}_1$, $\mathcal{A}_1 \mathcal{A}_2$, and $\mathcal{A}_2 \mathcal{A}_2$ in this population are 40, 40, and 20 respectively. These genotype frequencies imply that $(c_1, c_2)$ is proportional to $(4, 3)$. Table~\ref{ex1} shows how the average phenotypic change and $\lambda$ are affected by each step in a sequence of gene substitutions leading to an increase in $p$ but tending to keep $\lambda$ constant. The first column gives the gene substitution. In this sequence the two types of substitution alternate, but this is not an essential feature. The second column gives the numbers of the genotypes after the gene substitution. The third column gives the cumulative change in the total phenotypic measurements (the mean phenotype times the population size) divided by the number of gene substitutions. The fourth column gives the new value of $\lambda$ after the gene substitution.

\begin{table}
\caption{\emph{Sequence of experimental gene substitutions yielding the average effect.}}\label{ex1}
\begin{center}
\begin{tabular}{l l l l}
  \hline
  experimental change & genotype numbers & $\frac{\Delta (\mu N)}{\textrm{number of changes}}$ & $\lambda$ \\
  \hline
  --- & 40, 40, 20 & --- & 1/2 \\
  $\mathcal{A}_1 \mathcal{A}_1 \rightarrow \mathcal{A}_1 \mathcal{A}_2$ & 39, 41, 20 & 3 & .5387821 \\
  $\mathcal{A}_1 \mathcal{A}_2 \rightarrow \mathcal{A}_2 \mathcal{A}_2$ & 39, 40, 21 & 1/2 & .4884005 \\
  $\mathcal{A}_1 \mathcal{A}_1 \rightarrow \mathcal{A}_1 \mathcal{A}_2$ & 38, 41, 21 & 4/3 & .5266291 \\
  $\mathcal{A}_1 \mathcal{A}_2 \rightarrow \mathcal{A}_2 \mathcal{A}_2$ & 38, 40, 22 & 1/2 & .4784689 \\
  $\mathcal{A}_1 \mathcal{A}_1 \rightarrow \mathcal{A}_1 \mathcal{A}_2$ & 37, 41, 22 & 1 & .5162776 \\
  $\mathcal{A}_1 \mathcal{A}_2 \rightarrow \mathcal{A}_2 \mathcal{A}_2$ & 37, 40, 23 & 1/2 & .4700353 \\
  $\mathcal{A}_1 \mathcal{A}_1 \rightarrow \mathcal{A}_1 \mathcal{A}_2$ & 36, 41, 23 & 6/7 & .5075483 \\
  \hline
\end{tabular}
\end{center}
\end{table}

It is readily confirmed that the final value of $\lambda$ is the closest to the starting value of $1/2$ that can be achieved with 7 gene substitutions. If we take population size to infinity, we can make the discrepancy between the original and new values of $\lambda$ as small as we please. 

In the special case of genotype-environment independence considered so far, where equalities such as $\mathbb{E}(Y \,|\, \mathcal{A}_1 \mathcal{A}_1) = \mathbb{E}[ Y \,|\, do(\mathcal{A}_1 \mathcal{A}_1) ]$ always hold, Fisher's experimental and regression definitions of the average effect coincide for constant $\lambda$. In the example above, after assigning each genotype an expected phenotypic value consistent with the magnitudes of the experimental effects, it is easily verified that the slope in the least-squares regression of phenotypic value on $\mathcal{A}_2$ gene count is 6/7.

\section*{5. Gene-Environment Correlation and Interaction}

As a preliminary matter, we note that any variable along a causal path (in the sense of Wright and Pearl) from genotype to phenotype must not be counted as environmental. For example, if dairy consumption affects stature, it is tempting to regard dairy consumption as an environmental (non-genotypic) variable with respect to stature. But if genetic variation affects lactose tolerance and thus the amount of milk consumed, assigning the effect of dairy consumption on stature to the environment ignores the fact that the path \emph{genotype} $\rightarrow$ \emph{lactose tolerance} $\rightarrow$ \emph{dairy consumption} $\rightarrow$  \emph{stature} ultimately begins with a genetic variable. This subtlety may have been among the reasons why Fisher favored ``speaking of the residue as non-genetic, rather than environmental \ldots'' \citep[p.~260]{bennett:1983}

It is worth asking whether Fisher intended the average effect to be defined in the event that genotypic and environmental causes are either dependent or non-additive. In many places he certainly assumed or argued for independence and additivity \citep{fisher:1918,fisher:1941,fisher:1953,fisher:1970}, and it has been asserted that Fisher's biometrical theory is meaningless if these conditions are not met \citep[\emph{e.g.},][]{vetta:1980}.

As \citet{price:1972} has pointed out, Fisher's exposition in \emph{The Genetical Theory} leaves much to be desired. A close reading of this text and Fisher's other writings, however, turns up many reasons to suspect that Fisher regarded independence and additivity as reasonable specifications for certain demonstrations and not as strictly necessary conditions for the average effect to be defined.
\begin{enumerate}
\item In the discussion of the average effect in \emph{The Genetical Theory}, Fisher did not explicitly refer to his other work where he made special assumptions regarding the environment.

\item The average effect is a key concept in the FTNS, which Fisher regarded as an exact and rigorous statement. One would like to believe that Fisher, having been trained in mathematical physics, would not have compared the FTNS to the second law of thermodynamics if the FTNS depended on assumptions regarding the environment that must always be approximations at best.

\item We can read that ``[t]he genetic variance as here defined is only a portion of the variance determined genotypically, and this will differ from, and usually be somewhat less than, the total variance to be observed'' \citep[p.~34]{fisher:1930:gtns}. The genotypic variance is greater than the total variance only if ``good'' genotypes tend to be found in ``bad'' environments, and thus Fisher was clearly allowing for the possibility of dependence.

\item In a letter to J. A. Fraser Roberts, Fisher wrote that
\begin{quote}
[t]here is one point in which Hogben and his associates are riding for a fall, and that is in making a great song about the possible, but unproved, importance of non-linear interactions between hereditary and environmental factors\ldots. What they do not see is that we ordinarily count as genetic only such part of the genetic effect as may be included in a linear formula and that we make a present to the environmentalists of such variation due to the combined action of genetic and environmental factors as is not expressible in such a formula. \citep[p.~260]{bennett:1983}
\end{quote} 
These remarks clearly show that Fisher did not regard genotype-environment interaction as an obstacle to defining the average effect.
\end{enumerate}
Emboldened by this evidence regarding the intended generality of the average effect, we extend our treatment to encompass gene-environment correlation and interaction.

We first suppose that genotypic and environmental causes act additively but are not independent. Additivity means that the experimental effect of a gene substitution remains the same regardless of the environment in which the experiment is carried out; varying the environment simply raises or lowers the expected phenotypic values of all three genotypes by the same amount. For instance,
\begin{equation}\label{nointeract}
\Delta Y \,|\, \mathcal{A}_1 \mathcal{A}_1 \rightarrow \mathcal{A}_1 \mathcal{A}_{2},  \mathcal{E}_i =
\Delta Y \,|\, \mathcal{A}_1 \mathcal{A}_1 \rightarrow \mathcal{A}_1 \mathcal{A}_{2},  \mathcal{E}_j
\end{equation}
for any choice of environments $\mathcal{E}_i$ and $\mathcal{E}_j$. In this case all of the discussion in previous sections continues to apply \emph{except} for the equivalence of the experimental and regression average effects. If some genotypes are more frequently found in favorable environments for phenotypic development, then the regression of phenotypic value on gene count does not have a simple genetic interpretation. 

Non-additivity means that at least one equality of the kind in (\ref{nointeract}) does not hold. The precise magnitude of the expected change upon an experimental gene substitution now depends on some aspect of the environment that the manipulated zygote will experience between the onset of development and the time of measurement. This case is problematic because now a quantity such as $\Delta Y \,|\, \mathcal{A}_1 \mathcal{A}_1 \rightarrow \mathcal{A}_1 \mathcal{A}_2$ is not necessarily equal to $\Delta Y \,|\, \mathcal{A}_1 \mathcal{A}_2 \rightarrow \mathcal{A}_1 \mathcal{A}_1$, since the genotypes $\mathcal{A}_1 \mathcal{A}_1$ and $\mathcal{A}_1 \mathcal{A}_2$ may tend to be found in different environments. This difficulty can be overcome by redefining expressions such as $\Delta Y \,|\, \mathcal{A}_1 \mathcal{A}_1 \rightarrow \mathcal{A}_1 \mathcal{A}_2$ so that each symbolizes a difference between experimental treatments rather than a difference between a treatment and an unperturbed control group. For example, (\ref{change1}) would become
\begin{equation*}
\Delta Y \,|\, \mathcal{A}_1 \mathcal{A}_1 \rightarrow \mathcal{A}_1 \mathcal{A}_2 = \mathbb{E}[Y \,|\, do(\mathcal{A}_1 \mathcal{A}_2) ] - \mathbb{E}[Y \,|\, do(\mathcal{A}_1 \mathcal{A}_1) ]. 
\end{equation*}
Seeking an equivalent generalization that retains the interventional form of (\ref{change1}) and (\ref{change2}), however, sheds substantially greater light on the problem.

Before taking up the issue of gene-environment interaction, it is helpful to review Fisher's motivation for holding $\lambda$ constant as a means to address gene-gene interaction. In order to formulate the FTNS, Fisher wished to quantify the causal effect of changing allele frequency while holding the environment constant. In his view the way in which alleles combine to form genotypes, as parameterized by $\lambda$, should be regarded as part of the environment. Although this choice may initially seem eccentric, because fitness differences among genotypes will typically change both $p$ and $\lambda$, it becomes reasonable when we realize that $\lambda$ may also change as a result of extrinsic events such as the formation or dissolution of geographical hindrances to random mating. 

There is an analogy here to Fisher's analysis of covariance to separate the direct and indirect effects of a given experimental manipulation on a focal outcome. For instance, in an experiment to determine whether a given fertilizer affects the purity of sugar extracted from sugar-beets, the experimenter may already know that the fertilizer affects the weight of the beet roots, which in turn affects sugar purity \citep[pp.~283--284]{fisher:1970}. The experimenter may wish to know whether the fertilizer affects sugar purity through a direct causal path, \emph{fertilizer} $\rightarrow$ \emph{sugar purity}, distinct from the indirect path \emph{fertilizer} $\rightarrow$ \emph{root weight} $\rightarrow$ \emph{sugar purity}. In certain cases adjustment for root weight by analysis of covariance yields the target quantity: the amount by which sugar purity would change upon application of the fertilizer, if root weight could be experimentally clamped to the value that it would have obtained in the control condition. Similarly, while gene substitutions that are not deliberately balanced as in (\ref{alpha}) will typically change both $p$ and $\lambda$, we can still mathematically define an average effect stipulating that $\lambda$ remains clamped to a constant value. This point of view is similar to one expressed by \citet{okasha:2008}.

Once we regard any change in how alleles are arranged into genotypes as environmentally caused, it perhaps becomes obvious that we should regard certain changes in the allotment of genotypes to environments as such. After all, a redistribution among environments might lead to changes in the phenotypic means of the genotypes. Such changes in the genotype-phenotype mapping, when caused by extrinsic events such as climate change, are readily classified as environmental in nature. This consideration suggests that the gene substitutions defining the average effect in the presence of genotype-environment interaction should be balanced in such a way that the phenotypic means of the genotypes remain constant.

Since equalities such as $\mathbb{E}(Y \,|\, \mathcal{A}_1 \mathcal{A}_1) = \mathbb{E}[Y \,|\, do(\mathcal{A}_1 \mathcal{A}_1)]$ do not hold when genotypes and environments are also dependent, there is ambiguity in what is meant by holding constant the phenotypic means. We first consider holding constant the \emph{observed} means. If the environments interacting with genotypes can be classified discretely, then we can write an equation like
\begin{equation}
\mathbb{E}(Y \,|\, \mathcal{A}_1 \mathcal{A}_1) = \sum_i \mathbb{P}( \mathcal{E}_i \,|\, \mathcal{A}_1 \mathcal{A}_1) \mathbb{E}(Y \,|\, \mathcal{A}_1 \mathcal{A}_1, \mathcal{E}_i) 
\end{equation}
for each genotype. Because genotypes and environments exhaust all possible causes of phenotypic variation, $\mathbb{E}(Y \,|\, \mathcal{A}_1 \mathcal{A}_1, \mathcal{E}_i)$ is equivalent to $\mathbb{E}[Y \,|\, do(\mathcal{A}_1 \mathcal{A}_1), do(\mathcal{E}_i) ]$. In a sense even the expectation operator is unnecessary because $Y$ is a deterministic function when both genotype and environment are specified.

Constancy of observed means requires constancy of the conditional probabilities taking the form $\mathbb{P}( \mathcal{E}_i \,|\, \mathcal{A}_1 \mathcal{A}_1)$. A candidate definition for the average effect is then
\begin{multline*}
2\alpha dp = \mu[p + dp, \lambda, \mathbb{P}(\mathcal{E}_1 \,|\, \mathcal{A}_1 \mathcal{A}_1), \ldots, \mathbb{P}(\mathcal{E}_n \,|\, \mathcal{A}_2 \mathcal{A}_2)] \\
- \mu[p, \lambda, \mathbb{P}(\mathcal{E}_1 \,|\, \mathcal{A}_1 \mathcal{A}_1), \ldots, \mathbb{P}(\mathcal{E}_n \,|\, \mathcal{A}_2 \mathcal{A}_2)].
\end{multline*}
The problem with this candidate definition, however, is that it can lead to a nonzero average effect even if in each environment neither gene substitution has a causal effect. This is because preserving a genotype's conditional probabilities of being found in the various environments may require that some gene substitutions be accompanied by the placement of the manipulated organism in a different environment; the resulting change in phenotype may then be entirely the result of the environmental change.

If we instead consider holding constant the \emph{experimental} means, then we obtain
\begin{align}\label{constant_do}
\mathbb{E}[Y \,|\, do(\mathcal{A}_1 \mathcal{A}_1)] &= \sum_i \mathbb{P}[ \mathcal{E}_i \,|\, do(\mathcal{A}_1 \mathcal{A}_1)] \, \mathbb{E}[Y \,|\, do(\mathcal{A}_1 \mathcal{A}_1), \mathcal{E}_i] \nonumber\\
&= \sum_i \mathbb{P}( \mathcal{E}_i) \mathbb{E}(Y \,|\,\mathcal{A}_1 \mathcal{A}_1, \mathcal{E}_i). 
\end{align}
The left-hand side is the expected phenotypic value upon sampling a zygote at random and, if its genotype is not $\mathcal{A}_1 \mathcal{A}_1$, making it so. Since changing the genotype of a zygote cannot affect its environment, we have $\mathbb{P}[ \mathcal{E}_i \,|\, do(\mathcal{A}_1 \mathcal{A}_1)] = \mathbb{P}(\mathcal{E}_i)$ for each $i$ and thus a justification of the second line. Therefore preserving the experimental means only requires a constant marginal distribution of environmental states. Of course, we can always abide by this constraint if we never foster any manipulated organism in a different environment. This ensures that a nonzero average effect is indeed an average of genetic effects, at least one of which would turn out to be nonzero under experimental control.

Hence a natural definition of the average effect in the presence of genotype-environment interaction is
\begin{equation}\label{alpha_ge}
\alpha = \frac{ \sum_i c_{1,i} (\Delta Y \,|\, \mathcal{A}_1 \mathcal{A}_1 \rightarrow \mathcal{A}_1 \mathcal{A}_2, \mathcal{E}_i) + c_{2,i} (\Delta Y \,|\, \mathcal{A}_1 \mathcal{A}_2 \rightarrow \mathcal{A}_2 \mathcal{A}_2, \mathcal{E}_i) }
{ \sum_i c_{1,i} + c_{2,i} },
\end{equation}
where
\begin{align*}
c_{1,i} = c_1 \mathbb{P}( \mathcal{E}_i ), \\
c_{2,i} = c_2 \mathbb{P}( \mathcal{E}_i).
\end{align*}

\section*{6. Average Effects of Individual Alleles}

We will now explain how the experimental average effect of an individual allele may be defined for a locus with any number of alleles. Since there are ${n \choose 2}$ possible gene substitutions at a locus with $n$ alleles, we can no longer speak of a single average effect in the case of $n > 2$, and thus an extension of this kind is plainly necessary. In the second edition of \emph{The Genetical Theory}, we can read that ``[w]ith multiple allelomorphism it is convenient to define [the average effect of an allele] by the effect of substituting any chosen gene for a random selection of the genes homologous with it'' \citep[p.~35]{fisher:1958:gtns}. This definition can be explicated with respect to a given allele, say $\mathcal{A}_1$, as follows. Immediately after fertilization but before the onset of any developmental events, we select the allelic type of a gene to be changed into $\mathcal{A}_1$ in such a way that the probabilities of selection are equal to the allele frequencies. That is, if the vector of allele frequencies is $(p_1, \ldots, p_n)$, then the gene to be changed is $\mathcal{A}_1$ with probability $p_1$, $\mathcal{A}_2$ with probability $p_2$, and so on. If the gene to be changed happens to be $\mathcal{A}_1$ itself, then the $\mathcal{A}_1$ $\rightarrow$ $\mathcal{A}_1$ change will have no phenotypic consequence. For all changes other than the null change, the choice of the undisturbed gene in the genotype is made in such a way that the population's rules of genotype formation are preserved. If genotypes and environments are both dependent and interacting, then the marginal distribution of environmental states must be considered as in (\ref{alpha_ge}). The expected change in the phenotype of the manipulated organism is then $\alpha_1$, the average effect of $\mathcal{A}_1$.

From this definition we can derive some important consequences. Let $N_k$ stand for the number of $\mathcal{A}_k$ genes in the population. The total number of genes is $\sum_{k=1}^n N_k = N$. Among the $n$ experiments defining the individual average effects, choose one to perform with a probability equal to its corresponding allele frequency. The expected vector of allele frequencies following the randomly chosen experiment is then
\begin{equation}
\sum_{k=1}^n \frac{ N_k }{N } \left\{ \sum_{\ell=1}^n \frac{N_\ell }{N } \left[ \left( \frac{N_1}{N}, \ldots,  \frac{N_n}{N}  \right) + \frac{1}{N} \mathbf{e}_k - \frac{1}{N} \mathbf{e}_\ell \right]  \right\},
\end{equation}
where $\mathbf{e}_k$ is the vector of length $n$ with element unity at position $k$ and zeroes elsewhere.
After some algebra we find that the first element of the expected vector is $ N_1 \left(\sum N_k \right)^2/N^3 = p_1$, the second is $ N_2 \left(\sum N_k \right)^2/N^3 = p_2$, and so on. The expected outcome of the randomly chosen experiment is a population with exactly the same allele frequencies, rules of genotype formation, and phenotypic mean. We have thus proved that the experimental average effects satisfy
\begin{equation}\label{constraint}
\sum_{k=1}^n p_k \alpha_k = 0.
\end{equation}
With the generalization of the experimental average effect given in the next section, (\ref{constraint}) holds at any one of arbitrarily many multiallelic loci. In the case of a single locus, (\ref{constraint}) holds for the regression average effects in (\ref{ewens_regress}) \citep{ewens:2011}, and agreement of the regression and experimental average effects thus requires the mean subtraction in that expression.

Let us apply the definition of the individual average effect to the biallelic example in Table~\ref{ex1}. There are initially 120 $\mathcal{A}_2$ genes in this population of 200 total genes. If we perform the experiment defining $\alpha_1$, then with probability .40 the population gene numbers remain at $(80,120)$ and with probability .60 the numbers become $(81,119)$. In the event of a non-null substitution, with probability $4/7$ (given by $\frac{c_1}{c_1 + c_2}$) the change is $\mathcal{A}_1 \mathcal{A}_2$ $\rightarrow$ $\mathcal{A}_1 \mathcal{A}_1$ and with probability $3/7$ (given by $\frac{c_2}{c_1 + c_2}$) it is $\mathcal{A}_2 \mathcal{A}_2$ $\rightarrow$ $\mathcal{A}_1 \mathcal{A}_2$. The expected outcome of the experiment is thus a population with gene numbers $(80.6,119.4)$ and, up to the limits of finite size, the same value of $\lambda$. Using simple probability calculus, we can calculate that the numerical value of $\alpha_1$ is $-18/35$. 

In summary, the experiment defining $\alpha_1$ will lead to the null substitution $\mathcal{A}_1$ $\rightarrow$ $\mathcal{A}_1$ with probability $p_1$ (in which case the causal effect is zero) and to the substitution $\mathcal{A}_2$ $\rightarrow$ $\mathcal{A}_1$ with probability $p_2$ (in which case the effect is equal in magnitude to the average effect of gene substitution with respect to the entire locus). Therefore $\alpha_1$ must  be equal to $(p_1)(0) + (p_2)(-\alpha)$, and from this we can use $p_1 \alpha_1 + p_2 \alpha_2 = 0$ to derive $\alpha = \alpha_2 - \alpha_1$ algebraically. The meaning of this relation among the three average effects is as follows. The expected outcome of the experiment defining $\alpha_2$ is a population with gene numbers $(79.6,120.4)$ and nearly the same value of $\lambda$. Now suppose that we perform the ``opposite'' of the experiment defining $\alpha_1$, on average reducing the number of $\mathcal{A}_1$ genes rather than increasing them. We compose this experiment with the one defining $\mathcal{A}_2$, which in our example has a numerical value of $12/35$. The population is thus expected to proceed through the sequence $(80,120)$ $\rightarrow$ $(79.4,120.6)$ $\rightarrow$ $(79,121)$, preserving $\lambda$ at each step. The final state is precisely the one expected upon performing the experiment defining $\alpha$, the average effect of gene substitution for the entire locus. We can see in what sense the average effect of gene substitution ($6/7$) is equal to the effect of removing one gene ($18/35$) and then replacing it with another ($12/35$). 

\section*{7. Average Effects in the Case of Multiple Loci}

In the case of a single locus with two alleles, we can just as well define the average effect of gene substitution as
\begin{equation}\label{altdef}
\alpha = \frac{1}{2} \frac{ \partial \mu(p, \lambda) }{ \partial p },
\end{equation}
where $\mu$ is defined as in (\ref{exp_mean}). From this starting point, we can derive the equivalence of the regression (\ref{regress}) and experimental (\ref{alpha}) definitions in the case of genotype-environment independence.
(\ref{altdef}) fills the lacuna in Wright's casual use of the expression
\begin{equation*}
\frac{d \overline{W}}{dp},
\end{equation*}
to which \citet{fisher:1941} strongly objected. The explicit dependence of $\mu$ on $\lambda$, a measure of departure from random combination of genes, meets the criticism that ``the numerator involves the average of [the phenotype] for a number of different genotypes \ldots exceeding the number of gene frequencies $p$ on which their frequencies are taken to depend'' (p.~57).

It is interesting that the only genetic condition governing the gene substitutions defining the average effect for a single biallelic locus is the constancy of $\lambda$, a parameter that depends on the genotype frequencies but not the genotypic means. One might have thought that these means, appearing as they do in (\ref{regress}), must play some role in the weighting of the two possible gene substitutions. It is then natural to ask whether the generalization to multiple loci retains the appealing feature that constancy of appropriately quantified departures from Hardy-Weinberg and linkage disequilibrium is sufficient---without any additional information regarding the genotypic means---for an experimental average effect to agree with its corresponding partial regression coefficient. According to our analysis in the Appendix, the multilocus average effects do not in fact retain this feature. That is, we would like to define the multilocus average effect of allele $i_k$ at locus $k$, $\mathcal{A}^{(k)}_{i_k}$, as
\begin{equation}\label{baddef}
\alpha^{(k)}_{i_k} = \frac{1}{2} \frac{ \partial \mu(\mathbf{p}, \bm{\lambda}) }{ \partial p^{(k)}_{i_k} },
\end{equation}
where $\mathbf{p}$ is now a vector of allele frequencies at several loci, $p^{(k)}_{i_k}$ being the element corresponding to $\mathcal{A}^{(k)}_{i_k}$, and $\bm{\lambda}$ is a vector of whatever measures of departure from random combination are preserved under the appropriately balanced gene substitutions. However, as will be demonstrated, such a mean-invariant description of the average effects does not seem to exist.

To set up a weaker definition of the multilocus average effects, we require some additional definitions and notational conventions. Suppose that there are $L$ causal loci, in the sense of (\ref{defcause}), affecting the focal phenotype. Suppose also that there are $n_\ell$ alleles $\mathcal{A}^{(\ell)}_{i_\ell}$ $(i_\ell = 1, \ldots, n_\ell)$ at locus $\ell$. We have already stipulated that $p^{(\ell)}_{i_\ell}$ is the frequency of allele $\mathcal{A}^{(\ell)}_{i_\ell}$. Put $i = (i_1, \ldots, i_L)$ and denote the gamete $\mathcal{A}^{(1)}_{i_1} \cdots \mathcal{A}^{(L)}_{i_L}$ by the multi-index $i$. In addition, denote the frequency of the ordered multilocus genotype containing gametes $i$ and $j$ as $P_{ij}$. 

Define the \emph{coefficient of departure from random combination},
\begin{equation}\label{theta}
\theta_{ij} = \frac{ P_{ij} }{ \prod_k p^{(k)}_{i_k} p^{(k)}_{j_k} },
\end{equation}
as the ratio of the (ordered) whole-genome genotype $ij$ to the products of its constituent allele frequencies. The $\theta_{ij}$ are thus measures of both Hardy-Weinberg and linkage disequilibrium; they are all equal to unity if and only if the rules of genotype formation call for the random combination of all genes. Special cases of this coefficient were introduced by \citet{kimura:1958}, although \citet{nagylaki:1992} has pointed out that some of Kimura's expressions employing these coefficients are incorrect. To capture how the experimental gene substitutions defining the average effects change the departures from random combination, let
\begin{equation}
\mathring{\theta}_{ij} = \frac{ \Delta P_{ij} }{P_{ij} } - \sum_k \left( \frac{ \Delta p^{(k)}_{i_k} }{ p^{(k)}_{i_k} } +  \frac{ \Delta p^{(k)}_{j_k} }{ p^{(k)}_{j_k} } \right)
\end{equation}
denote the relative change in $\theta_{ij}$. In the limit of infinitesimal changes, this is equivalent to the logarithmic derivative of $\theta_{ij}$.

Now the experimenter must ascertain the mean of each whole-genome genotype by experimental control and then fit the equation
\begin{equation}\label{regress_manyloci}
\mathbb{E}[Y \,|\, do(ij)] = \mu +  \alpha_{ij} + \varepsilon_{ij}, \quad \textrm{where $\alpha_{ij} = \alpha_i + \alpha_j$, $\alpha_i = \sum_{k=1}^L \alpha^{(k)}_{i_k} $},
\end{equation}
to the treatment means thus obtained. The $\alpha^{(k)}_{i_k}$ are the average effects of the individual alleles. The residuals $\varepsilon_{ij}$ will reflect both dominance and epistasis, and in the general case it does not seem profitable to separate the two in the manner that \citet{kimura:1958} attempted. The fitting is accomplished by seeking the vector of average effects, $\bm{\alpha}$, that minimizes the sum of squares
\begin{equation}\label{gensumsq}
\sum_{i,j} P_{ij} \varepsilon_{ij}^2.
\end{equation}
Whereas the minimization defines the $\varepsilon_{ij}$ uniquely, the $\alpha^{(k)}_{i_k}$ are so far defined only up to a constant term in the sense that one constant may be added to the average effects at one locus and the same constant subtracted from the average effects at another locus without changing the minimum sum of squares \citep{ewens:2011}. The experimental average effect of a given allele, however, is obviously not defined only up to a constant term but rather must be equal to the precise number determined by the experiment of replacing a random homologous gene with a gene of the given allelic kind. In the Appendix we show that performing a non-null substitution in this experiment, in a manner preserving the rules of genotype formation, amounts to weighting the possible gene substitutions such that the scalar quantity
\begin{equation}\label{gencond}
\overline{\varepsilon \, \mathring{\theta}} = \sum_{i,j} P_{ij} \varepsilon_{ij} \mathring{\theta}_{ij}
\end{equation}
is equal to zero. Another way to phrase this key result is that the vanishing of $\overline{\varepsilon \, \mathring{\theta}}$ is a necessary and sufficient condition for the regression and experimental average effects to coincide in the case of genotype-environment independence. \citet{kimura:1958} showed that constancy of $\lambda$ suffices for $\overline{\varepsilon \, \mathring{\theta}}$ to vanish in the case of a single biallelic locus; it is worth mentioning that even in this simplest possible case there do not generally exist changes in the genotype frequencies such that each individual $\mathring{\theta}_{ij}$ vanishes.

Our theoretical experimenter can of course perform all $\sum_{k=1}^L n_k$ experiments to determine the unique values of the elements in the vector $\bm{\alpha}$. However, given our demonstration that the mean of the experimental average effects at any given locus is equal to zero, it suffices to impose (\ref{constraint}) for each locus as a constraint on the minimization of (\ref{gensumsq}). The proof of (\ref{constraint}) is still valid for each of multiple loci because the vanishing of $\overline{\varepsilon \, \mathring{\theta}}$ along each possible branch of the random experiment implies that the expected change in phenotypic mean must be equal to
\begin{equation}\label{constraint2}
2 \sum_{i_k}^{n_k}  \mathbb{E} \left( \Delta p^{(k)}_{i_k} \right) \alpha^{(k)}_{i_k},
\end{equation}
and since the expected outcome of the experiment is a population with the same allele frequencies, (\ref{constraint}) is assured.

The vanishing of $\overline{\varepsilon \, \mathring{\theta}}$ preserves the population's rules of genotype formation in the following sense. Although the number of parameters required to describe departure from random combination of genes increases very rapidly with the number of alleles and loci, (\ref{gencond}) implies it is not necessary for each and every such parameter to stay constant. It is enough, roughly speaking, for the average change in these parameters to equal zero. $\overline{\varepsilon \, \mathring{\theta}}$ is similar in form to the weighted average of the relative changes in the departures from random combination, those genotypes with large non-additive residuals being weighted more heavily.

The expression
\begin{equation}\label{def_manyloci}
\alpha^{(k)}_{i_k} = \frac{1}{2} \left( \frac{\partial }{\partial p^{(k)}_{i_k} }\sum_{i,j} P_{ij} \mathbb{E} [ Y \,|\, do(ij)] \right)_{ \overline{ \varepsilon \, \mathring{ \theta } } = 0 }
\end{equation}
may therefore serve as the definition of the experimental average effect in the case of multiple loci. 

Let us recapitulate the meaning of (\ref{def_manyloci}). Our variable of interest is the population average of the experimentally determined phenotypic means of the genotypes. If genotypes and environments are dependent, this variable is not the same as the population mean $\mathbb{E}(Y)$. Partial differentiation with respect to the frequency of allele $\mathcal{A}^{(k)}_{i_k}$ indicates that we examine how our variable of interest responds to the replacement of a small number of randomly chosen homologous genes with genes of the given allelic kind. The constraint on the partial derivative indicates that we consider only those counterfactual populations that can be reached from the original population by experimental replacements that result in the vanishing of (\ref{gencond}). The factor of $\frac{1}{2}$ is owed to diploidy.

It may seem from the form of the constrained derivative that this definition contains an element of circularity, since the $\varepsilon_{ij}$ are defined relative to the average effects in (\ref{regress_manyloci}). Any such concern should be dispelled by the fact that (\ref{def_manyloci}) fully encodes our argument from (\ref{regress_manyloci}) to (\ref{constraint2}), which provides an unambiguous sequence of instructions for the theoretical experimenter to follow. The Appendix provides some numerical examples.

\section*{8. Average Effects and Natural Selection}

At this point the reader may be questioning the need for defining the average effect in terms of causality, as might be revealed by experimentally controlled gene substitutions. Modern texts give only the regression definition \citep{lynch:1998,burger:2000}, and those who are accustomed to these accounts may resist the new notation and new way of thinking. 

We have already given one strong motivation to adopt the criterion of sensitivity to experimental manipulation: the need to distinguish a causal variant from the non-causal markers in LD with it. Another motivation is that dependence of genotypes and environments is a frequent occurrence. For instance, a major concern in GWAS is ensuring that discovered associations are not attributable to population stratification, which is essentially a form of confounding. A well-known apocryphal example is the ``chopstick gene.'' A geneticist performing a GWAS of chopstick skill in a large sample containing both Europeans and East Asians will undoubtedly find many marker loci failing to satisfy the equality
\begin{equation}
E(Y \,|\, \mathcal{A}_1 \mathcal{A}_1) = E(Y \,|\, \mathcal{A}_1 \mathcal{A}_2) = E(Y \,|\, \mathcal{A}_2 \mathcal{A}_2)
\end{equation}
even if, unbeknownst to the geneticist, the corresponding equality (\ref{defcause}) is obeyed at all loci linked to the statistically significant markers. This is because the Europeans and East Asians differ both in allele frequencies at these loci and in the prevalence of chopstick use; the latter difference presumably has arisen for reasons having nothing to do with genetics. A regression of the observed phenotypic values on gene count will nevertheless lead to a nonzero ``average effect'' in violation of both Fisher's verbal definition and common sense.

GWAS investigators attempt to control confounding by including all other genotyped markers in the regression. Since the number of genotyped markers typically exceeds the sample size, techniques such as principal components and mixed linear modeling are typically employed \citep{price:2006,zhou:2012}. The reason for the frequent effectiveness of these techniques is that genomic background become an extremely good proxy for the subpopulation to which a given sample member belongs as the number of loci grows large \citep{edwards:2003}. However, one can construct examples where partialing out other loci fails to deal with confounding \citep{mathieson:2012}, and in any case a theoretical definition whose usefulness depends on contingent quantities such as genome size and genetic diversity is inherently unattractive.

Perhaps the most conspicuous failure of the regression definition occurs in the very situation that motivated Fisher to define the average effect. This is when the phenotype is fitness itself. In this case the regression average effect will generically fail to be proportional to the partial change in genetic mean per change in allele frequency \emph{even if} the genotypic and environmental causes of fitness variation are additive and initially independent.

A simple simulation will bear out this perhaps surprising claim. The simulated organism follows a life cycle consisting of non-overlapping generations. The population size is 20,000. Fitness is determined by a single locus and the environment; the frequency of $\mathcal{A}_2$ is initially 1/2, and the population mated at random in the previous generation. The genotypic fitnesses---the values of $\mathbb{E}[Y \,|\, do(\mathcal{A}_1 \mathcal{A}_1)]$, $\mathbb{E}[Y \,|\, do(\mathcal{A}_1 \mathcal{A}_2)]$, $\mathbb{E}[Y \,|\, do(\mathcal{A}_2 \mathcal{A}_2)]$---are .4, .5, and .6 respectively. We determine the phenotypic fitness of each individual in the following way. Immediately after fertilization but before the onset of viability selection, an environmental disturbance of .3 in absolute value is added to each individual's genotypic fitness. Positive and negative disturbances are equally probable. This scheme ensures that genotypes and environments are independent at this time.

Whether an individual withstands viability selection to mate with a random fellow survivor is determined by a discrete approximation of an exponential process. We stipulate ten discrete time intervals between fertilization and reproduction, each of which an individual survives with a probability chosen so that the probability of surviving all ten intervals is equal to the individual's phenotypic fitness. By dividing the time between fertilization and mating into more intervals, we could more closely approach a true continuous-time model, where the logarithm of phenotypic fitness would be similar to the Malthusian parameter. Ten intervals, however, suffice to make the point at issue.

\begin{table}
\caption{\emph{Evolutionary change across time intervals in a simulated organism.}}\label{ex2}
\begin{center}
\begin{tabular}{l c c c}
  \hline
   & $\beta$ & $\Delta p$ & $\Delta A$ \\
  \hline
  fertilization & .100 & $9.33 \times 10^{-3}$ & $1.87 \times 10^{-3}$ \\
  time 1 & .091 & $8.07 \times 10^{-3}$ & $1.61 \times 10^{-3}$ \\
  time 2 & .084 & $5.86 \times 10^{-3}$ & $1.17 \times 10^{-3}$ \\
  time 3 & .079 & $6.11 \times 10^{-3}$ & $1.22 \times 10^{-3}$ \\
  time 4 & .073 & $5.08 \times 10^{-3}$ & $1.02 \times 10^{-3}$ \\
  time 5 & .069 & $4.72 \times 10^{-3}$ & $9.45 \times 10^{-4}$ \\
  time 6 & .065 & $4.10 \times 10^{-3}$ & $8.20 \times 10^{-4}$ \\
  time 7 & .062 & $3.47 \times 10^{-3}$ & $6.95 \times 10^{-4}$ \\
  time 8 & .060 & $3.00 \times 10^{-3}$ & $5.91 \times 10^{-4}$ \\
  time 9 & .059 & $1.28 \times 10^{-3}$ & $2.57 \times 10^{-4}$ \\
  time 10 & .060 & --- & --- \\
  \hline
\end{tabular}
\end{center}
\end{table}

Table~\ref{ex2} shows the evolution of this population from fertilization to mating. The first column gives the time interval. The second column gives the regression average effect---the slope in the regression of phenotypic values on $\mathcal{A}_2$ gene count among those individuals alive at the beginning of the time interval; $\beta$ is the conventional notation for a regression coefficient. The third column gives the change in $\mathcal{A}_2$ frequency from the beginning of the current time interval to the beginning of the next. The fourth column gives the change in the mean genotypic fitness from the beginning of the current time interval to the beginning of the next. Because the effect of substituting $\mathcal{A}_2$ for $\mathcal{A}_1$ does not depend on the allelic type of the undisturbed gene, the experimental average effect is of course .10. In this case of additive gene action, the genotypic value is the same as the ``breeding'' or ``additive genetic'' value, which is now often denoted by the symbol $A$. 

Immediately after fertilization, the regression and experimental average effects coincide, as expected from the fact that genetic values and environmental disturbances are initially independent. The change in mean genetic value from fertilization to the beginning of the first time interval is equal to two times the experimental average effect times the change in allele frequency. The relation $\Delta A = 2 \alpha \Delta p$ in fact holds for each transition from one time interval to the next. The relation $\Delta A = 2 \beta \Delta p$, however, does not hold for any transition besides the first. Note the decline in $\beta$, far greater and more systematic than can be explained by sampling fluctuations, with the passage of time.

What explains the increasing discrepancy between $\alpha$ and $\beta$? This is an example of what some methodologists call \emph{selection bias} \citep[\emph{e.g.},][]{pearl:2009}. Suppose that intelligence and athletic ability are uncorrelated in the population at large. However, if we limit our observations to the students attending a university that uses both of these attributes as admissions criteria, then we will find that intelligence and athleticism are negatively correlated. If we learn that a student at this university is academically undistinguished, then it becomes more probable that the student is a good athlete. Otherwise the student would likely not have been admitted. 

Similarly, if there is some relation between fitness at different points of the lifespan, then with the passage of time the genetic and environmental causes of fitness will tend to become correlated even if they were initially independent. If we learn that a particular survivor of a rigorous selection scheme has an unfit genotype, then it becomes more probable that the organism has benefited from a favorable environment. This same principle explains why selection tends to induce deviations from Hardy-Weinberg and linkage equilibrium \citep{bulmer:1980,nagylaki:1992,burger:2000,ewens:2004}; if we find that a survivor has an unfit gene at one genomic position, it becomes more probable that the survivor bears fit genes at other positions. As stated previously, the dependence of genotypes and environment leads to a divergence between the experimental and regression average effects, and the latter then has no straightforward genetic interpretation.

It is important to note that our example does not necessarily impugn the validity of the FTNS, under the regression definition of the average effect, with respect to organisms living in discrete time. This is because in this model the FTNS has come to be interpreted as concerning the change in mean breeding value between generations, and the correctness of the FTNS is preserved when the mean is measured upon fertilization and the regression average effect is measured at the beginning of the parental generation. However, because our model places deaths along a temporal dimension between birth and mating, it should properly be classified as a continuous-time model. The FTNS is intended to apply at every point in continuous time, and therefore our argument for the experimental definition of the average effect retains its full force for organisms following such a life cycle.

Fisher knew that selection bias with respect to the outcome variable prevents regression coefficients from being interpretable. In \emph{Statistical Methods for Research Workers}, he pointed out that the application of a selection process to the outcome variable will change the regression line \citep[p.~130]{fisher:1970}. It is thus rather curious that Fisher never mentioned this principle in connection with natural selection, a form of selection bias that is always and everywhere operating. 

The regression definition is made viable by stipulating the use of ``true'' or ``intrinsic'' phenotypic measurements as the outcome variable rather than the actual measurements. This approach, which we adopt in the Appendix, may be natural and inevitable in the case of multiple loci. Because of the need to know the residuals in the multilocus case, it does not seem possible to banish the concept of least-squares linear regression from the theory of average effects. The concepts of regression and causality need to work together. Needless to say, the notion of causality remains an essential partner in this collaboration. A definition calling for the regression of ``true'' phenotypic measurements on gene content really amounts to replacing the observed phenotypic means of the three genotypes in (\ref{regress}) with the experimental means, which requires the same $do$ operator incorporated in (\ref{alpha}) and (\ref{alpha_ge}). The instance of $do$ in (\ref{def_manyloci}) actually covers two points where we must invoke experimental control: once in the determination of the genotypic means, breeding values, and non-additive residuals, and again in the replacement of randomly chosen homologous genes to resolve the non-uniqueness of the individual average effects. To capture what Fisher intended by the average effect in a formal and transparent way, we cannot easily avoid a special notation for singling out causal relations from merely correlational ones. 

\section*{9. Discussion}

\citet{falconer:1985} had the good sense to intuit that sensitivity to physical change was important to Fisher's conception of the average effect. Indeed, among all twentieth-century scientists, Fisher might have been the one most likely to incorporate the distinction between an observed excess and a causal effect into a formal theory. The discrepancy that Falconer thought he had uncovered between Fisher's regression and experimental definitions of the average effect can be reconciled, in the case of genotype-environment independence, by using a specific weighted average of the two possible gene substitutions rather than a naive average. If the phenotype is affected by one biallelic locus, the weights are chosen so that a population subject to gene substitutions in numbers proportional to the weights retains the same value of $\lambda = Q^2/(PR)$, a parameter describing the way in which alleles are combined into genotypes. If genotypes and environments interact non-additively, then the gene substitutions must also be balanced with respect to the marginal distribution of environmental states. This balancing has the desirable property of preserving the experimentally ascertained phenotypic means of the genotypes. In the case of multiple loci, there is no longer a fixed parameterization of genotype formation to which the weightings of the gene substitutions must conform, but in a loose sense the changes in the departures from random combination must average out to zero. These restrictions are requirements for a change in allele frequency ``without change in the environment, or in the mating system [rules of genotype combination].'' When genotypes and environments are dependent---which must always be the case, even if only slightly, as a result of natural selection---the experimental definition is to be preferred.


\citet{fisher:1941} gave one reason why a definition based on experimental gene substitutions may be inferior to one based on passive observations of a static population (although later in this paper he reverted to the language of gene substitutions). He pointed out that changes in the frequencies of the different genotypes may feed back to change the phenotypic means themselves. He gave the example of experimental gene substitutions increasing milk yield, which lead to females in the next generation who can leverage their superior nourishment to provide even more milk to their own offspring. Fisher wished to discount such knock-on effects---presumably because they are too complex to form general rules about them. These knock-on effects can be positive or negative. When fitnesses are frequency-dependent, the knock-on effects of naturally selected changes in allele frequencies can steadily decrease the mean fitness of the population \citep{nowak:2006}. The approach of a female-skewed sex ratio to a stable fifty-fifty equilibrium in a polygynous species can be an example of precisely this phenomenon (\cites[pp.~141-143]{fisher:1930:gtns}[p.~232]{bennett:1983}). Therefore Fisher consigned changes in the genotype-phenotype mapping---the $\mathbb{E}[Y \,|\, do(ij)]$---brought about by gene substitutions with all other possible such changes, including those brought about by unpredictable changes in climate, predators, parasites, and so on. Our preferred resolution of the dilemma raised by the cascade of additional phenotypic changes that may be initiated by a physical gene substitution is to stipulate the constancy of (\ref{change1}) and (\ref{change2}), for instance, in the experimental definition of the average effect. That is, the average effect is calculated on the assumption that the prevailing genotype-phenotype mapping will not itself change as a result of the gene substitutions. This is equivalent to the \emph{stable unit treatment value assumption} (SUTVA) in the Neyman-Rubin counterfactual framework. 

SUTVA may often have a reasonable interpretation. For example, in the cases of fecundity selection and frequency-dependent fitnesses of game-theoretic strategies, we may interpret each causal effect as the expected phenotypic change upon placing a manipulated organism in a virtual environment containing the same mixture of types constituting the undisturbed population. In any event finding an interpretation of SUTVA may not be important in most biological situations, so long as any frequency-dependent changes ensuing from the experimental manipulation of a few individuals can be neglected in a theoretically infinite population.

It is the constancy of the $\mathbb{E}[Y \,|\, do(ij)]$ rather than the constancy of the corresponding observed phenotypic means that is satisfied by the gene substitutions defining the average effect in the case of genotype-environment dependence and interaction. This striking fact further affirms the priority of causal quantities over observables that may have no causal interpretation.

A renewed understanding of the average effect is especially timely given the enablement of GWAS by modern technology and the upsurge of research into the inheritance of fitness in human populations \citep{stearns:2010}. The findings of the \citet{encode:2012} indicate that the fine-mapping of the variants with nonzero experimental average effects responsible for a given association signal may turn out to be less onerous than was once supposed. However, care is needed as researchers isolate variants with ever smaller average effects, which will be difficult to distinguish from spurious signals generated by subtle confounding or selection bias.

An appealing feature of GWAS is the availability of a complementary study design, pioneered by \citet{spielman:1993}, that offers nearly the entirety of the benefits inhering in experimental control. According to Mendel's laws, a parent passes on a randomly chosen gene from each of its homologous pairs to a given offspring. Given the applicability of Mendel's laws, we can then treat the genotype of an offspring given the parental genotypes much like a treatment in a randomized experiment. It follows that a significant association between transmission of a particular allele and the focal phenotype cannot be the result of confounding; in the absence of selection bias, the only feasible explanation is linkage with a locus where the average effect is nonzero. Fisher himself noted this feature of family-based studies:
\begin{quote}
Genetics is indeed in a peculiarly favoured condition in that Providence has shielded the geneticist from many of the difficulties of a reliably controlled comparison. The different genotypes possible from the same mating have been beautifully randomized by the meiotic process. A more perfect control of conditions is scarcely possible, than that of different genotypes appearing in the same litter. \citep[p.~7]{fisher:1952}
\end{quote}
Family-based studies have successfully been used to replicate findings from studies of nominally unrelated individuals \citep{langoallen:2010,turchin:2012}, and this is another way in which the thought experiments defining the average effect are becoming less like \emph{Gedanken} and more like routine empirical operations. We note that when \citet{spielman:1993} introduced their family-based test, their null hypothesis was no linkage with a causal locus despite the presence of population association. This test and its variants have since often been used to test the null hypothesis that there is neither linkage nor association. We anticipate that there will be a trend back toward the original form of the test. Because parent-offspring trios and sets of siblings can be difficult to recruit and require more genotyping, investigators find it convenient to test for population association in large samples of unrelated individuals. Those markers showing evidence of association can then be interrogated, however, for linkage with loci where there are nonzero average effects. The follow-up cohorts of families will typically be much smaller and less likely to yield genome-wide significant $p$-values, but it will be reasonable to require less stringent evidence or merely overall sign agreement greatly exceeding 50 percent. This procedure can provide a check on whether the association stage is producing an acceptably low rate of false positives with respect to the causal hypothesis of a nonzero average effect---which, of course, is not strictly the same as the statistical hypothesis of a nonzero partial regression coefficient.

We note that family-based studies are not immune to selection bias intervening between fertilization and the time of measurement, which may rise to an appreciable level in studies of phenotypes strongly affecting fitness. This may be a challenge for gene-trait mapping studies conducted in the near future.

It may be tempting to define the average effect in terms of a hypothetical family-based study. However, whereas rejecting the null hypothesis of a zero average effect requires only the assumptions of Mendel's laws, effect estimation requires additional assumptions and thus does not seem particularly suited for a theoretical definition after all \citep{ewens:2008}. 

Finally, we comment on the role of the average effect in the FTNS. We write the breeding (additive genetic) value of a given individual as
\begin{equation}\label{breedval}
A = \sum_{\ell =1}^L \sum_{i_\ell=1}^{n_{\ell}} \chi \left( \mathcal{A}^{(\ell)}_{i_\ell} \right) \alpha^{(\ell)}_{i_\ell},
\end{equation}
where $\chi(\cdot)$ is a function giving the number of $\mathcal{A}^{(\ell)}_{i_\ell}$ genes (0, 1, or 2) present in the individual's genotype. The variance in breeding values, $\textrm{Var}(A)$, is now called the \emph{additive genetic variance}, and the ratio $\textrm{Var}(A)/\textrm{Var}(Y)$ the \emph{heritability in the narrow sense}. It is important to keep in mind that these breeding values are linear functions of \emph{experimental} average effects; we are building up a predicted value for a given individual from the causal effects of the genes present in the genotype.

The FTNS states that the partial change in mean fitness attributable to changes in allele frequencies caused by natural selection is proportional to the additive genetic variance in fitness, which can be shown to equal
\begin{equation}\label{partchange}
2 \sum_{\ell =1}^L \sum_{i_\ell=1}^{n_{\ell}} p^{(\ell)}_{i_\ell} a^{(\ell)}_{i_\ell} \alpha^{(\ell)}_{i_\ell}, 
\end{equation}
where the meaning of $a^{(\ell)}_{i_\ell}$ is as follows. If genotypes and environments are independent, then this quantity is the average excess of $\mathcal{A}^{(\ell)}_{i_\ell}$, which is usually defined as the difference in mean fitness between the bearers of the given allele and the entire population. (\ref{partchange}) is invariably derived under the assumption that genotypes and environments are independent. Because under our definitions the values of the experimental average effects do not depend on the extent of genotype-environment dependence, it follows that the breeding values and hence the additive genetic variance are also insensitive to genotype-environment dependence. The equality of (\ref{partchange}) with $\textrm{Var}(A)$ is thus fully valid in our account---given the following modification regarding $a^{(\ell)}_{i_\ell}$.

If genotypes and environments are not independent, $a^{(\ell)}_{i_\ell}$ in (\ref{partchange}) is not exactly the same as the average excess defined by \citet[p.~35]{fisher:1958:gtns}. It is rather the average excess that \emph{would} be observed if genotypes were distributed randomly among environments. In other words each $a^{(\ell)}_{i_\ell}$ only reflects confounding with other genetic loci and not with environmental causes. To repeat, this is a consequence of the fact that our experimental average effects---and hence all quantities derived from them, including the additive genetic variance---are sensitive only to the marginal distribution of environmental states. Every factor in (\ref{partchange}), including the $a^{(\ell)}_{i_\ell}$, must therefore be equal to whatever they would be under genotype-environment independence, the standard setting in which (\ref{partchange}) is calculated. If the ``full'' average excesses were substituted into (\ref{partchange}), then the expression would no longer be interpretable as a variance; it could then possibly be negative. 

It is well known that the change in the frequency of $\mathcal{A}^{(\ell)}_{i_\ell}$ is proportional to the product of $p^{(\ell)}_{i_\ell}$ and the actual difference in mean fitness between the bearers of the given allele and the entire population \citep[\emph{e.g.},][]{price:1970}. From the fact that the difference is not necessarily equal to our $a^{(\ell)}_{i_\ell}$, we learn that there is partition of the total change in allele frequency between the change caused by natural selection and the change attributable to how genotypes are distributed across environments varying in severity. This partition is in the same spirit of Fisher's conditions discussed previously. Like changes in the rules of genotype formation or the $\mathbb{E}[Y \,|\, do(ij)]$, deviations from genotype-environment independence cannot generally lead to an increase in fitness, and indeed the example set out in Table~\ref{ex2} demonstrates that the dependence induced by natural selection itself tends to retard the frequency increase of the superior allele.

Each increment of naturally selected change in allele frequency is a direct cause of a change in the mean fitness equal to $2 \alpha^{(\ell)}_{i_\ell}$. Any discrepancy between the total change and this partial change, summed over all loci and alleles, is owed to indirect effects acting through changes in the rules of genotype formation, the distribution of environmental states, or some other determinant of fitness. This completes the FTNS: the \emph{increase} in the mean fitness of a population caused exclusively by the effect of natural selection on allele frequencies---setting aside those \emph{changes} in fitness (which can be positive or negative) ascribable to other causes---is equal to the additive genetic variance in fitness.



Fisher's contributions to biology and applied mathematics were of course numerous and profound. Judging from his writing in \emph{The Genetical Theory}, however, we surmise that he considered the FTNS to be the most important of his achievements. The FTNS quantifies Darwin's notion of hereditary variation in fitness leading to adaptation and provides a deeper understanding of it. It is interesting that (\ref{partchange}), Fisher's ``supreme law of the biological sciences,'' explicitly encodes a distinction between an observed excess and a causal effect, the same distinction that animated his work on experimental design, which \citet{neyman:1967} praised as the greatest of Fisher's contributions to statistics. The FTNS was thus another blow struck by Fisher against his scientific adversary Karl Pearson, who believed it was possible both to study evolution mathematically and to discard the notion of causality. If causality appears inevitably in the formulation of a phenomenon as fundamental as evolution by natural selection, then it surely cannot be a dispensable ``fetish amidst the inscrutable arcana of modern science'' \citep[p.~xii]{pearson:1911}.


\section*{Acknowledgements}

We thank A. W. F. Edwards for sharing his unpublished work and his correspondence with the late Douglas Falconer, Sabin Lessard for helpfully answering one of our queries, and the reviewers for comments and suggestions that have greatly improved this paper. This work was supported by the Intramural Research Program of the NIDDK, NIH.

\section*{Appendix}

\renewcommand{\theequation}{A\arabic{equation}}
\setcounter{equation}{0}

\renewcommand{\thetable}{A\arabic{table}}
\setcounter{table}{0}

Here we explicitly derive the conditions under which the regression and experimental definitions of average effect are equivalent. We assume that the equivalence can always be secured in a meaningful way, either because genotypes and environments are independent or because the regression has been performed on the experimental genotypic means rather than the observed genotypic means. We will often refer to an experimental average effect in the sense of an arbitrary linear combination of relevant causal effects (differences between genotypic means) and narrow down our reference to particular linear combinations as the given argument proceeds. We first treat the case of a single biallelic locus, which is of special interest because it is possible here to find explicit expressions for the weights $c_1$ and $c_2$ in (\ref{alpha}).

Let $i$ stand for $\textrm{E}[ Y \,|\, do(\mathcal{A}_1 \mathcal{A}_1)]$, $j$ for $\textrm{E}[ Y \,|\, do(\mathcal{A}_1 \mathcal{A}_2)]$, and $k$ for $\textrm{E}[ Y \,|\, do(\mathcal{A}_2 \mathcal{A}_2)]$. This notation is similar to that of \citet{fisher:1918,fisher:1941}. By using the $do$ symbol, however, our argument below is meaningful even if genotypes and environments are dependent and non-additive.

To minimize the sum of squares
\begin{equation*}
P ( i - \nu + \alpha )^2 +
2Q ( j - \nu )^2 +
R [ k - \nu - \alpha ]^2,
\end{equation*}
we take partial derivatives with respect to $\nu$ and $\alpha$ and set them equal to zero. Solving the two resulting equations gives
\begin{equation}\label{alpha2}
\alpha = \frac{P(Q+R)(j - i) + R(P+Q)(k - j)}{PQ + QR + 2PR},
\end{equation}
which can easily be recognized as equivalent to (\ref{alpha}) in the case that genotypes and environments act additively. Using (\ref{constant_do}) to expand each experimental mean, we find that the numerator of (\ref{alpha2}) becomes
\begin{multline}
c_1 \left[ \sum_i \textrm{Pr}( \mathcal{E}_i) \textrm{E}(Y \,|\,\mathcal{A}_1 \mathcal{A}_2, \mathcal{E}_i) - \textrm{Pr}( \mathcal{E}_i) \textrm{E}(Y \,|\,\mathcal{A}_1 \mathcal{A}_1, \mathcal{E}_i) \right] \\
+ c_2 \left[ \sum_i \textrm{Pr}( \mathcal{E}_i) \textrm{E}(Y \,|\,\mathcal{A}_2 \mathcal{A}_2, \mathcal{E}_i) - \textrm{Pr}( \mathcal{E}_i) \textrm{E}(Y \,|\,\mathcal{A}_1 \mathcal{A}_2, \mathcal{E}_i)  \right],
\end{multline}
which means that (\ref{alpha2}) is also equivalent to (\ref{alpha_ge}).



Now consider the change in the mean phenotype caused by experimental gene substitutions.  The contribution to the population mean phenotype by the experimental means of the genotypes is given by
\begin{equation}
\mu = i P + 2jQ + k R,
\end{equation}
and the change in the population mean upon effecting the gene substitutions is
\begin{equation}\label{total_change}
d\mu = i dP + 2j dQ +k dR.
\end{equation}
%

The changes $dP, dQ, dR$ have two degrees of freedom. To express the changes in terms of a single change $dp$, we must obtain another condition, which can be expressed without loss of generality as $f(P,Q,R) = 0$. \citet{fisher:1941} gave the condition that $\lambda = Q^2/(PR)$ remains constant, but his concise argument has puzzled many commentators.


It turns out that Fisher set $d\mu = i dP + 2j dQ + k dR$ equal to $2 \alpha dp$ and equated the coefficients of $i, j, k$ \citep{edwards:1967}, which yields
\begin{align}
dP &= -2P(Q+R)dp/S,\nonumber \\
dQ &= Q(P-R)dp/S,\nonumber\\
dR &= 2R(P+Q)dp/S,
\label{conds}
\end{align}
where $S = P(Q+R)+R(P+Q)$. The function $f$ satisfies the differential equation
\begin{equation}
\frac{\partial f}{\partial P} dP +\frac{\partial f}{\partial Q}dQ +\frac{\partial f}{\partial R} dR =0.
\label{fcond}
\end{equation}
Inserting (\ref{conds}) into (\ref{fcond}) gives
\begin{equation}
-2P(Q+R)\frac{\partial f}{\partial P} +Q(P-R)\frac{\partial f}{\partial Q} +2R(P+Q)\frac{\partial f}{\partial R} =0.
\end{equation}
Now $(-2P(Q+R),Q(P-R),2R(P+Q))$ and $(\frac{\partial f}{\partial p},\frac{\partial f}{\partial Q},\frac{\partial f}{\partial R})$ can be regarded as two orthogonal vectors in three-space.  We want the second condition to be independent of the conservation of probability condition and not to be the trivial zero vector.  By inspection, we see that a solution is given by
\begin{align}
\frac{\partial f}{\partial P} &= \frac{\phi}{P},\nonumber\\
\frac{\partial f}{\partial Q} &= \frac{-2\phi}{Q},\nonumber\\
\frac{\partial f}{\partial R} &= \frac{\phi}{R},
\label{fconds}
\end{align}
where $\phi$ is an arbitrary function of $P,Q,R$.  A simple solution is given by setting $\phi$ equal to the constant $a$, whereupon (\ref{fconds}) can be integrated to obtain
\begin{equation}
f=-2a\ln Q + a\ln P + a\ln R  + a\ln \lambda,
\end{equation}
which gives the condition $Q^{2a}=(\lambda PR)^{a}$.  $a=1$ gives the condition expressed in terms of the classic Fisher parameter. Conversely, if we let $\phi=PRQ^{-2}$ then we get $f=PRQ^{-2}-(1/\lambda)$, which also gives the Fisher parameter.  

Taking the partial second derivatives gives the compatibility conditions that $\phi$ must satisfy:
\begin{align}
\frac{1}{P}\frac{\partial\phi}{\partial R}&=\frac{1}{R}\frac{\partial \phi}{\partial P}, \nonumber\\
\frac{1}{P}\frac{\partial\phi}{\partial Q}&=\frac{-2}{Q}\frac{\partial \phi}{\partial P}, \nonumber\\
\frac{1}{R}\frac{\partial\phi}{\partial Q}&=\frac{-2}{Q}\frac{\partial \phi}{\partial R}.
\end{align}
Hence, any differentiable function of $PRQ^{-2}$ is a solution.  This then implies that $f$ can be any differentiable function of $PRQ^{-2}$ as well.  This shows that the average phenotypic increment caused by a number of experimental gene substitutions is the same as the slope in the regression of the phenotype on the experimental genotypic means if  the substitutions are performed in a background where any function of $PRQ^{-2}$ is held constant, with $\lambda$ being the simplest one.

We now treat a phenotype affected by an arbitrary number of multiallelic loci. As shown in Section 7, the experimentally determined phenotypic means of the whole-genome genotypes can be expressed as
\begin{equation*}
\mathbb{E} [Y \,|\, do(ij)] = \mu + \alpha_{ij} + \varepsilon_{ij}.
\end{equation*}
In the remainder we abbreviate $\mathbb{E} [Y \,|\, do(ij)]$ as $G_{ij}$ and set $a_{ij}=G_{ij} - \mu$, which obeys the condition
$\sum_{i,j} P_{ij} a_{ij} =0$.  

The average effects can be written as $\alpha_{ij}=\alpha_i +\alpha_j=\sum_\ell(\alpha_{i_\ell}^{(\ell)} + \alpha_{j_\ell}^{(\ell)})$ and are obtained by minimizing
\begin{equation}
\sum_{i,j}P_{ij}(a_{ij}-\alpha_{ij})^2.
\end{equation}
The minimum obeys the condition
\begin{equation}
p_{i_k}^{(k)} a_{i_k}^{(k)} = \sum_{i\nmid i_k}\sum_j P_{ij} \alpha_{ij} = \sum_{i\nmid i_k}\sum_j P_{ij}\sum_\ell \left( \alpha_{i_\ell}^{(\ell)} + \alpha_{j_\ell}^{(\ell)} \right),
\label{regressionsolution}
\end{equation}
where
\begin{equation}
p_{i_k}^{(k)}a_{i_k}^{(k)} =  \sum_{i\nmid i_k}\sum_ja_{ij}P_{ij}
\label{avgexcessdef}
\end{equation}
defines the average excesses. A sum running over $i \nmid i_k$ should be understood as a sum over all multi-indices $i$ where the $k$th element is fixed to $i_k$. These relations imply that $\sum_{ij}P_{ij} \alpha_{ij}=0$, which also implies that $\sum_{ij}P_{ij}\varepsilon_{ij} =0$.

Equation (\ref{regressionsolution}) can be rewritten as
\begin{multline}
p_{i_k}^{(k)} a_{i_k}^{(k)}= \left( p_{i_k}^{(k)}+q_{i_ki_k}^{(kk)} \right) \alpha_{i_k}^{(k)} + \sum_{\ell\ne k}\sum_{i_\ell}p_{i_k i_\ell}^{(k \ell)}\alpha_{i_\ell}^{(\ell)}  + \sum_\ell \sum_{j_\ell\ne i_k} q_{i_k j_\ell}^{(k \ell)}\alpha_{j_\ell}^{(\ell)} \\
\equiv \sum_{j_\ell} H^{(k \ell)}_{i_k j_\ell} \alpha_{j_\ell}^{(\ell)},
\label{statdef}
\end{multline}
where 
\begin{equation*}
p^{(k\ell)}_{i_k i_\ell} = \sum_{i \nmid i_k, i_\ell} \sum_j P_{ij}
\end{equation*}
denotes the frequency of gametes that carry $\mathcal{A}^{(k)}_{i_k}$ and $\mathcal{A}^{(\ell)}_{i_\ell}$ and
\begin{equation*}
q^{(k \ell)}_{i_k j_\ell} = \sum_{i \nmid i_k} \sum_{j \nmid j_\ell} P_{ij}
\end{equation*}
denotes the frequency of all multilocus genotypes that carry $\mathcal{A}^{(k)}_{i_k}$ and $\mathcal{A}^{(\ell)}_{j_\ell}$ on different chromosomes. The matrix $\mathsf{H}$ in (\ref{statdef}) is constructed as follows. Let $\mathbf{p}$ denote the vector of allele frequencies, $\mathbf{a}$ the vector of average excesses, and $\bm{\alpha}$ the vector of average effects. These vectors have length $\sum_k^L n_k$, and their elements are ordered by locus. We can then define
\begin{equation}
\mathsf{H} = \mathsf{D} + \mathsf{P} + \mathsf{Q},
\end{equation}
where $\mathsf{D}$ is the diagonal matrix with the components of $\mathbf{p}$ on the diagonal, $\mathsf{P}$ is the matrix with entries $p^{(k\ell)}_{i_k i_\ell}$ if $k \neq \ell$ and 0 otherwise, and $\mathsf{Q}$ is the matrix with entries $q^{(k \ell)}_{i_k j_\ell}$ \citep{ewens:1992,castilloux:1995}. 
We will use the notation $\mathbf{p} \cdot \mathbf{a}$ to designate the component-wise product of the vectors $\mathbf{p}$ and $\mathbf{a}$, i.e., $(\mathbf{p} \cdot \mathbf{a})_i = \mathbf{p}_i \mathbf{a}_i$. (\ref{statdef}) can thus be rewritten again as
\begin{equation}
\bm{\alpha} = \mathsf{H}^{-1} (\mathbf{p} \cdot \mathbf{a})
\end{equation}
subject to suitable constraints on $\bm{\alpha}$. We will shortly see that these constraints turn out to be (\ref{constraint}) for each locus. Given our ordering convention, the element $H^{(k \ell)}_{i_k j_\ell}$ lies in the row of $\mathsf{H}$ corresponding to allele $\mathcal{A}^{(k)}_{i_k}$ and the column corresponding to $\mathcal{A}^{(\ell)}_{i_\ell}$.

The total change in $\mu$ is
\begin{multline}
d\mu = \sum_{i_k}\sum_{i,j} G_{ij} \frac{\partial P_{ij}}{\partial p^{(k)}_{i_k}} dp_{i_k}^{(k)} = \sum_{i_k}\sum_{i,j} (\mu+\alpha_{ij} +\varepsilon_{ij})\frac{\partial P_{ij}}{\partial p^{(k)}_{i_k}} dp_{i_k}^{(k)} \\
= \sum_{i_k}\sum_{i,j} (\alpha_{ij} +\varepsilon_{ij})\frac{\partial P_{ij}}{\partial p^{(k)}_{i_k}} dp_{i_k}^{(k)}
\end{multline}
upon performing a number of experimental gene substitutions at locus $k$. Agreement of the experimental and regression average effects implies that this change must equal the change predictable from the breeding values,
\begin{equation}
d\mu = \sum_{i_k}\sum_{i,j} \alpha_{ij} \frac{\partial P_{ij}}{\partial p^{(k)}_{i_k}} dp_{i_k}^{(k)},
\end{equation}
which implies in turn that
\begin{equation}\label{fullcondition}
\sum_{i,j} \sum_{i_k} \varepsilon_{ij} \frac{\partial P_{ij}}{\partial p^{(k)}_{i_k}} dp_{i_k}^{(k)}=0
\end{equation}
is a necessary and sufficient condition for the experimental and regression average effects to coincide. The bald statement that the changes in genotype frequencies must somehow nullify the non-additive residuals, however, is not very revealing. We can render (\ref{fullcondition}) into a more insightful form by noting that
\begin{equation}
\sum_{i,j} P_{ij}\sum_{\ell=1}^L \left( \frac{\partial p_{i_\ell}}{p_{i_\ell} } +  \frac{\partial p_{j_\ell}}{p_{j_\ell} }\right)\varepsilon_{ij}=0
\end{equation}
because the sum over $\ell$ is a constant determined by the experimenter.
Using this, from (\ref{fullcondition}) we obtain
\begin{equation}
\sum_{ij} P_{ij}\left[ \frac{1}{P_{ij}} \frac{\partial P_{ij}}{\partial p^{(k)}_{i_k}} -\sum_\ell\left( \frac{\partial p_{i_\ell}}{p_{i_\ell} } +  \frac{\partial p_{j_\ell}}{p_{j_\ell} }\right)\right]\varepsilon_{ij}=0,
\label{nlcondition}
\end{equation}
which leads to (\ref{gencond}). This argument, which simplifies one given by \citet{lessard:1997}, can be used to construct a variety of quantities measuring departures from random combination. The $\theta_{ij}$ appear to be the simplest such quantities.

The criterion (\ref{nlcondition}) does not pick out a unique weighting of the possible gene substitutions for a given genetic architecture. It would be of great significance if a subset of the possible weights could be characterized in a manner that does not depend on the non-additive residuals. We have done this for a single biallelic locus, where the subset contains the singleton weighting of the two possible gene substitutions that conserves $\lambda$. If a general procedure for constructing such a residual-free characterization for any number of loci exists, then the following argument should be able to find it.


The contribution of the experimental genotypic means to the population mean is
\begin{equation}
\mu = \sum_{i,j} G_{ij} P_{ij}. 
\end{equation}
The definition of the experimental average effect can be written as
\begin{equation}
\alpha_{i_k}^{(k)} = \frac{1}{2}\frac{\partial \mu}{\partial p_{i_k}^{(k)}}.
\label{aepd}
\end{equation}
Imposing constancy of the experimental means, we can write the change in the population mean due to a change in frequency of allele $\mathcal{A}^{(k)}_{i_k}$ as
\begin{equation}
 \frac{\partial}{\partial p^{(k)}_{i_k}} \mu = \sum_{i,j} G_{ij} \frac{\partial P_{ij}}{\partial p^{(k)}_{i_k}} = \sum_{i,j} (G_{ij}-\mu) \frac{\partial P_{ij}}{\partial p^{(k)}_{i_k}}=\sum_{i,j} a_{ij}  \frac{\partial P_{ij}}{\partial p^{(k)}_{i_k}}, 
 \label{causaldef}
\end{equation}
using the fact that $\sum_{i,j}  \frac{\partial}{\partial p^{(k)}_{i_k}}P_{ij} = 0$.  The indeterminacy in the partial derivatives with respect to allele frequency will be resolved by the properties of $\bm{\lambda}$ in (\ref{baddef}) that emerge from the subsequent analysis.

Substituting (\ref{causaldef}) and (\ref{aepd}) into (\ref{statdef}) using (\ref{avgexcessdef}) gives the condition
\begin{equation}
 \sum_{i\nmid i_k}\sum_ja_{ij}P_{ij}=  \frac{1}{2}\sum_j \sum_{j_\ell} H^{(k \ell)}_{i_k j_\ell}\sum_{m,n} a_{mn}  \frac{\partial P_{mn}}{\partial p^{(\ell)}_{j_\ell}} 
\label{condition}
\end{equation}
for each $i_k$ and $k$. Closed-form solutions of these partial differential equations will not exist in general.  However, using symmetry conditions and properties of $\mathsf{H}$, we may infer some necessary conditions on the genotype frequencies that must be satisfied. 




%
%

We first note that the image space of $\mathsf{H}$ contains all permissible vectors of allele-frequency changes \citep{lessard:1995}. Since $\mathsf{H}$ is invertible on its image space, we may operate on (\ref{condition}) by the inverse of $\mathsf{H}$ (which we call $\mathsf{J}$) and thereby separate the PDE system into a set of $\sum n_\ell$ ordinary differential equations, which we denote by
\begin{equation}
2\sum_{i_k} J_{j_k i_k} p_{i_k}^{(k)} a_{i_k}^{(k)} = \sum_{m,n} a_{mn} \frac{\partial P_{mn}}{\partial p^{(k)}_{j_k}}.
\label{newcond}
\end{equation}
We may now select any row of (\ref{newcond}), expand the $p_{i_k}^{(k)} a_{i_k}^{(k)}$ in terms of the $a_{mn}$, and equate the LHS and RHS coefficients of $a_{mn}$.  This will result in a set of $\prod n_\ell \times \prod n_\ell$ ordinary differential equations of the form
\begin{equation}
\frac{1}{P_{mn}}\frac{\partial P_{mn}}{\partial p^{(k)}_{j_k}} = \phi_{mn}(j_k),
\label{ODEcond}
\end{equation}
where $\phi_{mn}(j_k)$ is some linear combination of the elements of the vector $J_{j_\ell=j_k, i_\ell}$. From this point the $a_{mn} = \alpha_{mn} + \varepsilon_{mn}$ no longer appear in the argument, and it follows that we must be finding properties of a solution that depends on neither the breeding values nor the non-additive residuals.

Conserved quantities imposed by (\ref{condition}), which can be used to form elements of $\bm{\lambda}$, can be constructed by taking linear combinations of the ODEs such that 
\begin{equation}
\sum_{m,n} \sigma_{mn} \frac{1}{P_{mn}}\frac{\partial P_{mn}}{\partial p^{(k)}_{j_k}} = 0,
\end{equation}
from which we obtain conserved measures of departure from random combination assuming the form
\begin{equation}
\frac{\prod_{\{\sigma>0\}} P_{\alpha\beta}}{\prod _{\{\sigma<0\}}P_{\gamma\delta}} = \lambda_\sigma,
\end{equation}
where $\sigma_{mn}$ is some set of coefficients that are positive, zero, or negative.  These conserved quantities will form a set of necessary conditions for the equivalence of the experimental and regression definitions of the average effects.

Note that the coefficients of $a_{mn}$ on the LHS of (\ref{newcond}) are grouped according to the $a_{i_k}^{(k)}$.  Thus all of the $a_{mn}$ expressed in a given $a_{i_k}^{(k)}$ will have the same coefficient (one of the elements of $\mathsf{J}$). We can thus construct conserved measures of Hardy-Weinberg and linkage disequilibrium without an explicit calculation of $\mathsf{J}$ because we know which sets of coefficients are equal.  

Our first numerical example is of a single locus with three alleles (Table~\ref{one_triallelic}). The case of a single locus with any number of alleles was analytically treated by \citet{kempthorne:1957}. The equating of coefficients along the $i$th row of (\ref{newcond}) leads to the matrix of equations
\begin{equation}\label{triallelic_cond}
\begin{pmatrix}
J_{i1} = \dfrac{1}{P_{11}} \dfrac{ \partial P_{11} }{ \partial p_i } & J_{i1} = \dfrac{1}{P_{12}} \dfrac{ \partial P_{12} }{ \partial p_i } & J_{i2} = \dfrac{1}{P_{21}} \dfrac{ \partial P_{21} }{ \partial p_i } \\
 & & \\
J_{i2} = \dfrac{1}{P_{22}} \dfrac{ \partial P_{22} }{ \partial p_i } & J_{i1} = \dfrac{1}{P_{13}} \dfrac{ \partial P_{13} }{ \partial p_i } & J_{i3} = \dfrac{1}{P_{31}} \dfrac{ \partial P_{31} }{ \partial p_i } \\
 & & \\
 J_{i3} = \dfrac{1}{P_{33}} \dfrac{ \partial P_{33} }{ \partial p_i } & J_{i2} = \dfrac{1}{P_{23}} \dfrac{ \partial P_{23} }{ \partial p_i } & J_{i3} = \dfrac{1}{P_{32}} \dfrac{ \partial P_{32} }{ \partial p_i } \\
\end{pmatrix}
\end{equation}
for allele $i$. 
The notation $P_{ij}$ now means the ordered genotype with alleles $i$ and $j$. 
This matrix gives a set of nine conditions plus conservation of probability that must be satisfied to ensure the equality of (\ref{condition}).  However, given that there are only six unique genotypes, these conditions are overdetermined and will not necessarily be solvable.  We can attempt to formulate a solvable set by combining these conditions.
We can see that the second and third elements in a given row of this matrix must equal the sum of the elements in the first column corresponding to the homozygous bearers of the relevant alleles. For example,
\begin{equation}
\frac{1}{P_{12}} \frac{ \partial P_{12} }{ \partial p_i } +  \frac{1}{P_{21}} \frac{ \partial P_{21} }{ \partial p_i } = \frac{1}{P_{11}} \frac{ \partial P_{11} }{ \partial p_i } +  \frac{1}{P_{22}} \frac{ \partial P_{22} }{ \partial p_i } = J_{i1} + J_{i2},
\end{equation}
and these equations lead collectively to the three conserved measures of Hardy-Weinberg disequilibrium
\begin{equation}\label{triallelic_lambda}
\lambda_{12} = \frac{ P_{12}^2 }{P_{11} P_{22}}, \quad \lambda_{13} = \frac{ P_{13}^2 }{P_{11} P_{33}}, \quad \lambda_{23} = \frac{ P_{23}^2 }{P_{22} P_{33}}.
\end{equation}
Two of the allele frequencies and these three conserved quantities appear to be a complete specification of the six genotype frequencies. By the implicit function theorem, invertibility of the Jacobian at any solution ($p_1$, $p_2$, $\lambda_{12}$, $\lambda_{13}$, $\lambda_{23}$) specifying a valid vector of genotype frequencies ensures that there are unique solutions for small perturbations of the allele frequencies. Numerical testing suggests that invertibility of the Jacobian is a generic property of this five-dimensional system.

\begin{table}
\caption{A trait affected by a single triallelic locus.} \label{one_triallelic}
\begin{center}
\begin{tabular}{l c c r}
  \hline
  genotype & $\mathbb{E}[Y \,|\, do(\cdot)]$ & frequency & $\varepsilon$ \\
  \hline
  $\mathcal{A}_1 \mathcal{A}_1$ & 10 & .2 & $-.3402778$ \\
  $\mathcal{A}_2 \mathcal{A}_2$ & 13 & .2 & .2152778 \\
  $\mathcal{A}_3 \mathcal{A}_3$ & 12 & .2 & $-.6875$ \\
  $\mathcal{A}_1 \mathcal{A}_2$ & 11 & .2 & $-.5625$ \\
  $\mathcal{A}_1 \mathcal{A}_3$ & 14 & .1 & 2.4861111 \\
  $\mathcal{A}_2 \mathcal{A}_3$ & 13 & .1 & .2638889 \\
  \hline
\end{tabular}
\end{center}
\end{table}

Given the numerical values in Table~\ref{one_triallelic}, what is the experimental average effect of substituting $\mathcal{A}_2$ for $\mathcal{A}_1$? There are three ways in which this gene substitution can be brought about: $\mathcal{A}_1 \mathcal{A}_1$ $\rightarrow$ $\mathcal{A}_1 \mathcal{A}_2$, $\mathcal{A}_1 \mathcal{A}_2$ $\rightarrow$ $\mathcal{A}_2 \mathcal{A}_2$, and $\mathcal{A}_1 \mathcal{A}_3$ $\rightarrow$ $\mathcal{A}_2 \mathcal{A}_3$. The causal effects of these three substitutions are 1, 2, and $-1$ respectively. 

We first attempt to satisfy the weaker criterion that (\ref{gencond}) is equal to zero by determining which weighted average of the first two substitutions yields the smallest absolute value of $\overline{ \varepsilon \, \mathring{\theta}  }$. To calculate a discrete approximation of the $\mathring{\theta}_{ij}$, we use a population size of 10,000. We examine all integer weights such that the weights sum to 90. There are 91 such weighted averages, and it turns out that the weights $(70,20)$ yield the minimum. In fact, the absolute value of $\overline{ \varepsilon \, \mathring{\theta}  }$ yielded by these weights is roughly $1.5 \times 10^{-16}$, which is nearly within machine error of zero. The 90 other weighted averages lead to absolute values of $\overline{ \varepsilon \, \mathring{\theta}  }$ exceeding $1 \times 10^{-4}$.

These weights lead to an experimental average effect, $\alpha_2 - \alpha_1$, equaling 11/9. In the case of a single locus, the regression average effects (which we now denote by $\beta$) do not require the imposition of (\ref{constraint}) to be identified, and the calculations yielding the values of the $\varepsilon_{ij}$ in Table~\ref{one_triallelic} also give us ($-0.7798611$, 0.4423611, 0.39375) as the numerical value of ($\beta_1$, $\beta_2$, $\beta_3$). It appears that $\beta_2 - \beta_1$ is exactly equal to 11/9.

We can use a different pair of substitutions, say $\mathcal{A}_1 \mathcal{A}_2$ $\rightarrow$ $\mathcal{A}_2 \mathcal{A}_2$ and $\mathcal{A}_1 \mathcal{A}_3$ $\rightarrow$ $\mathcal{A}_2 \mathcal{A}_3$, to yield the experimental average effect $\alpha_2 - \alpha_1$. We examine all integer weightings of these two substitutions such that the weights sum to 270. It turns out that the weighting $(200,70)$ yields the minimum. The absolute value of $\overline{ \varepsilon \, \mathring{\theta}  }$ yielded by these weights is roughly $4 \times 10^{-16}$, again nearly within machine error of zero, whereas the 270 other weighted averages all lead to absolute values of $\overline{ \varepsilon \, \mathring{\theta}  }$ exceeding $3 \times 10^{-4}$. These minimizing weights again lead to an experimental average effect of 11/9. It is rather interesting that the neighboring weights $(199,71)$ and $(201,69)$ lead to such higher values of $\overline{ \varepsilon \, \mathring{\theta}  }$ despite the numerical closeness of these weighted averages and the fineness of our discretization. In fact, we have chosen to present this example because of this phenomenon, which we conjecture to be related to the fact that the $\alpha_2 - \alpha_1$ happens to be rational and thus exactly equal to some integer-weighted average of the causal effects.

Evidently it should not be possible to obtain a valid average effect by using only the substitutions $\mathcal{A}_1 \mathcal{A}_1$  $\rightarrow$ $\mathcal{A}_1 \mathcal{A}_2$ and $\mathcal{A}_1 \mathcal{A}_3$ $\rightarrow$ $\mathcal{A}_2 \mathcal{A}_3$. Examining all integer weights summing to 1000, we find that $\overline{ \varepsilon \, \mathring{\theta}  }$ declines linearly from $(0,1000)$ to $(1000,0)$; the absolute minimum of $\overline{ \varepsilon \, \mathring{\theta}  }$ is thus attained at a boundary, and it is not especially small ($\sim 2 \times 10^{-2}$).

We examine whether our conception of individual average effects is valid. Using the method of minimizing $\overline{ \varepsilon \, \mathring{\theta}  }$, we find that $\alpha_2 - \alpha_3$ is approximately .049. According to our notion of substituting $\mathcal{A}_2$ for a random homologous gene, $\alpha_2$ must be equal to $p_1(\alpha_2 - \alpha_1) + p_3(\alpha_2 - \alpha_3)$. In our example ($p_1$, $p_2$, $p_3$) happens to be (.35, .35, .30), which leads to $.4425$ as the approximate numerical value of $\alpha_2$. This is in good agreement with $\beta_2$. Continuing this exercise, we can satisfy ourselves that ($\alpha_1$, $\alpha_2$, $\alpha_3$) and ($\beta_1$, $\beta_2$, $\beta_3$) are equal.

We now attempt to satisfy the stronger criterion that the quantities in (\ref{triallelic_lambda}) remain constant. The numerical value of ($p_1$, $p_2$, $\lambda_{12}$, $\lambda_{13}$, $\lambda_{23}$) is (35/100, 35/100, 1/4, 1/16, 1/16), and a perturbation of ($-1/1000$, $1/1000$, 0, 0, 0) leads to a numerical solution that specifies another valid vector of genotype frequencies. The weighting of the possible gene substitutions satisfying the changes in genotype frequencies is typically not unique. In a population of size $10^8$, one permissible vector of weights for our example can be reasonably well approximated by
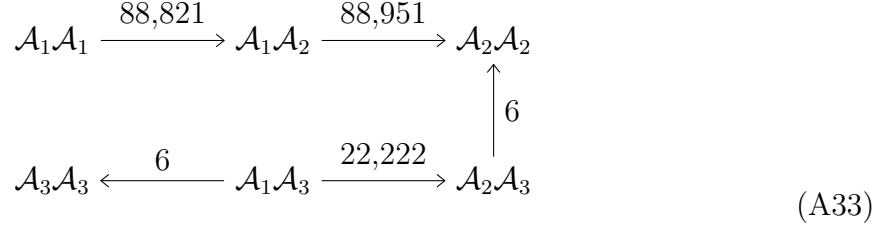
\begin{equation}
\begin{tikzpicture}
  \matrix (m) [matrix of math nodes,row sep=3em,column sep=4em,minimum width=2em] {
     \mathcal{A}_1 \mathcal{A}_1 & \mathcal{A}_1 \mathcal{A}_2 & \mathcal{A}_2 \mathcal{A}_2 \\
     \mathcal{A}_3 \mathcal{A}_3 & \mathcal{A}_1 \mathcal{A}_3 & \mathcal{A}_2 \mathcal{A}_3 \\};
  \path[-biggertip]
    (m-1-1) edge node [above] {88,821} (m-1-2)
    (m-1-2) edge node [above] {88,951} (m-1-3)
    (m-2-2) edge node [above] {6} (m-2-1) 
    (m-2-2) edge node [above] {22,222} (m-2-3)
    (m-2-3) edge node [right] {6} (m-1-3) ;
\end{tikzpicture}    
\end{equation}
where the label of each arrow indicates how many gene substitutions of that kind are to be performed. Notice that there are 12 gene substitutions involving a genotype containing the allele $\mathcal{A}_3$. For each $\mathcal{A}_3$ gene created by $\mathcal{A}_1 \mathcal{A}_3$ $\rightarrow$ $\mathcal{A}_3 \mathcal{A}_3$, another $\mathcal{A}_3$ is destroyed by $\mathcal{A}_2 \mathcal{A}_3$ $\rightarrow$ $\mathcal{A}_2 \mathcal{A}_2$, and the net result is the same frequency of $\mathcal{A}_3$. These 12 substitutions turn out to be a way of decreasing the number of $\mathcal{A}_1$ genes and increasing the number of $\mathcal{A}_2$ without directly converting one to the other. We might as well pair each $\mathcal{A}_1 \mathcal{A}_3$ $\rightarrow$ $\mathcal{A}_3 \mathcal{A}_3$ with $\mathcal{A}_2 \mathcal{A}_3$ $\rightarrow$ $\mathcal{A}_2 \mathcal{A}_2$, treating each such pair as a single substitution. The weighted average of the gene substitutions is then
\begin{equation*}
\frac{ 88,821(1) + 88,951(2) + 22,222(-1) + 6(-2 + 0) }{88,821 + 88,951 + 22,222 + 6  },
\end{equation*}
which diverges from 11/9 at the fourth decimal place.

We now apply our argument to the case of two biallelic loci. Here we will encounter a contradiction.

The equating of coefficients along the row of (\ref{newcond}) corresponding to allele $\mathcal{A}^{(k)}_{i_k}$ now leads to the matrix of equations
\begin{equation}\label{two_biallelic_cond}
\hspace*{-.8cm}
\begin{pmatrix}
J_{i1} + J_{i3} = \frac{1}{P_{11,11}} \frac{ \partial P_{11,11} }{ \partial p^{(k)}_{i_k}} & J_{i1} + J_{i4} = \frac{1}{P_{12,11}} \frac{ \partial P_{12,11} }{ \partial p^{(k)}_{i_k} } & J_{i2} + J_{i3} = \frac{1}{P_{21,11}} \frac{ \partial P_{21,11} }{ \partial p^{(k)}_{i_k} } & J_{i2} + J_{i4} = \frac{1}{P_{22,11}} \frac{ \partial P_{22,11} }{ \partial p^{(k)}_{i_k} } \\
J_{i1} + J_{i3} = \frac{1}{P_{11,12}} \frac{ \partial P_{11,12} }{ \partial p^{(k)}_{i_k} } & J_{i1} + J_{i4} = \frac{1}{P_{12,12}} \frac{ \partial P_{12,12} }{ \partial p^{(k)}_{i_k} } & J_{i2} + J_{i3} = \frac{1}{P_{21,12}} \frac{ \partial P_{21,12} }{ \partial p^{(k)}_{i_k} } & J_{i2} + J_{i4} = \frac{1}{P_{22,12}} \frac{ \partial P_{22,12} }{ \partial p^{(k)}_{i_k} } \\
J_{i1} + J_{i3} = \frac{1}{P_{11,21}} \frac{ \partial P_{11,21} }{ \partial p^{(k)}_{i_k} } & J_{i1} + J_{i4} = \frac{1}{P_{12,21}} \frac{ \partial P_{12,21} }{ \partial p^{(k)}_{i_k} } & J_{i2} + J_{i3} = \frac{1}{P_{21,21}} \frac{ \partial P_{21,21} }{ \partial p^{(k)}_{i_k} } & J_{i2} + J_{i4} = \frac{1}{P_{22,21}} \frac{ \partial P_{22,21} }{ \partial p^{(k)}_{i_k} } \\
J_{i1} + J_{i3} = \frac{1}{P_{11,22}} \frac{ \partial P_{11,22} }{ \partial p^{(k)}_{i_k} } & J_{i1} + J_{i4} = \frac{1}{P_{12,22}} \frac{ \partial P_{12,22} }{ \partial p^{(k)}_{i_k} } & J_{i2} + J_{i3} = \frac{1}{P_{21,22}} \frac{ \partial P_{21,22} }{ \partial p^{(k)}_{i_k} } & J_{i2} + J_{i4} = \frac{1}{P_{22,22}} \frac{ \partial P_{22,22} }{ \partial p^{(k)}_{i_k} } \\
\end{pmatrix}
\end{equation}
plus conservation of probability that must be satisfied to ensure the equality of (\ref{condition}). An argument analogous to the one below (\ref{triallelic_cond}) shows that six quantities of the form
\begin{equation}
\lambda_{ij} = \frac{ P_{ij}^2 }{P_{ii} P_{jj}}
\end{equation}
must be conserved. If we do not assume that the double heterozygotes are phenotypically equivalent, then these six measures of Hardy-Weinberg disequilibrium, the allele frequencies at the two loci, and conservation of probability leave one more condition to specify ten genotype frequencies. 

Rearrange each element of (\ref{two_biallelic_cond}) to put the genotype frequency on one side and form the four column sums. Each such sum is the marginal frequency of a gamete. For example, we have
\begin{equation}
P_{11} = (J_{i1} + J_{i3})^{-1} \sum_{11,j} \frac{ \partial P_{11,j} }{ \partial p^{(k)}_{i_k} },
\end{equation}
which implies that
\begin{equation}
J_{i1} + J_{i3} = \frac{1}{P_{11}} \frac{ \partial P_{11} }{ \partial p^{(k)}_{i_k} }.
\end{equation}
Combining all columns, we get
\begin{equation}
\frac{ \partial P_{11} }{ \partial p^{(k)}_{i_k} } + \frac{ \partial P_{22} }{ \partial p^{(k)}_{i_k} } - \frac{ \partial P_{12} }{ \partial p^{(k)}_{i_k} } - \frac{ \partial P_{21} }{ \partial p^{(k)}_{i_k} } = 0,
\end{equation}
which yields the condition that
\begin{equation}
\zeta = \frac{P_{11}P_{22} }{P_{12} P_{21}  }
\end{equation}
remains constant. $\zeta$ is the measure introduced by \citet{kimura:1965}, and the multi-index notation immediately reveals that it is equal to unity in linkage equilibrium. 

The equality of the regression and experimental average effects for constant $\bm{\lambda} = (\lambda_{11,21}, \ldots, \lambda_{12,22}, \zeta)$ appears to conflict with the result of \citet{nagylaki:1976} that the stipulation of $\Delta \zeta = 0$ and random mating to reset the $\lambda_{ij}$ to unities among zygotes does not lead to the change in the mean phenotype equaling the summed products of average effects and changes in allele frequencies (in the case that the phenotype is fitness). Our next numerical example shows that we have indeed reached a contradiction (Table~\ref{two_biallelic}).

\begin{table}
\caption{A trait affected by two biallelic loci.} \label{two_biallelic}
\begin{center}
\begin{tabular}{l c c r}
  \hline
   genotype & $\mathbb{E}[Y \,|\, do(\cdot)]$ & frequency & $\varepsilon$ \\
  \hline
  $\mathcal{A}^{(1)}_1 \mathcal{A}^{(2)}_1 / \mathcal{A}^{(1)}_1 \mathcal{A}^{(2)}_1$ & 17 & .054 & 5.0100265 \\
  $\mathcal{A}^{(1)}_1 \mathcal{A}^{(2)}_2 / \mathcal{A}^{(1)}_1 \mathcal{A}^{(2)}_1$ & 12 & .036 & $-1.438691$ \\
  $\mathcal{A}^{(1)}_1 \mathcal{A}^{(2)}_2 / \mathcal{A}^{(1)}_1 \mathcal{A}^{(2)}_2$ & 13 & .257 & $-.8874187$ \\
  $\mathcal{A}^{(1)}_1 \mathcal{A}^{(2)}_1 / \mathcal{A}^{(1)}_2 \mathcal{A}^{(2)}_1$ & 14 & .140 & $-.3345667$ \\
  $\mathcal{A}^{(1)}_1 \mathcal{A}^{(2)}_2 / \mathcal{A}^{(1)}_2 \mathcal{A}^{(2)}_1$ & 18 & .080 & $-.7832893$ \\
  $\mathcal{A}^{(1)}_1 \mathcal{A}^{(2)}_1 / \mathcal{A}^{(1)}_2 \mathcal{A}^{(2)}_2$ & 10 & .039 & $-4.7832893$ \\
  $\mathcal{A}^{(1)}_1 \mathcal{A}^{(2)}_2 / \mathcal{A}^{(1)}_2 \mathcal{A}^{(2)}_2$ & 16 & .066 & 4.7679882 \\
  $\mathcal{A}^{(1)}_2 \mathcal{A}^{(2)}_1 / \mathcal{A}^{(1)}_2 \mathcal{A}^{(2)}_1$ & 15 & .041 & $-.6791599$ \\
  $\mathcal{A}^{(1)}_2 \mathcal{A}^{(2)}_1 / \mathcal{A}^{(1)}_2 \mathcal{A}^{(2)}_2$ & 11 & .029 & $-3.2178824$ \\
  $\mathcal{A}^{(1)}_2 \mathcal{A}^{(2)}_2 / \mathcal{A}^{(1)}_2 \mathcal{A}^{(2)}_2$ & 20 & .258 & .4233950 \\
  \hline
\end{tabular}
\end{center}
\end{table}

Numerical testing suggests that invertibility of the Jacobian is also a generic property of the nine-dimensional system ($p^{(1)}$, $p^{(2)}$, $\lambda_{11,21}$, \ldots, $\lambda_{12,22}$, $\zeta$). We numerically update the vector of genotype frequencies in Table~\ref{two_biallelic} by increasing the frequency of allele $\mathcal{A}^{(1)}_2$ by $10^{-6}$. The regression average effect at locus 1, as determined by the Levenberg-Marquardt algorithm, is approximately 2.4934. However, when we multiply this by two times $10^{-6}$, the result does not closely agree with $ G_{ij} \Delta P_{ij} $. The discrepancy is close to 12 percent and does not diminish as $\Delta p^{(1)}$ is made smaller. We conclude that we have falsified our initial assumption that a residual-free description of the average effects always exists.

Sampling vectors of initial genotype frequencies from the Dirichlet distribution, we find that the changes implied by constancy of $\bm{\lambda}$ in the case of two biallelic loci do not typically produce such a large discrepancy. The error is usually less than 7 percent. This suggests to us that there may exist a subset of weights, distinguished by the changes in the departures from random combination all being ``small'' in some sense, that can be mathematically described. We leave this issue to future research.

The vanishing of $\overline{ \varepsilon \, \mathring{\theta} }$ is still an applicable criterion. For example, the genotype $\mathcal{A}^{(1)}_1 \mathcal{A}^{(2)}_2$/ $\mathcal{A}^{(1)}_1 \mathcal{A}^{(2)}_1$ can be transformed into either double heterozygote, depending on whether the left or right gene at locus 1 is the target of the substitution. In one case the causal effect is 6, and in the other it is $-2$. Among all integer weightings of these two substitutions summing to 1000, the weights (562, 438) yield the minimum. The corresponding weighted average of the causal effects, $\alpha^{(1)}_2 - \alpha^{(1)}_1$, equals 2.496 and is also the closest to $\beta^{(1)}_2 - \beta^{(1)}_1 \approx 2.493$ that can be obtained given our discretization. The replacement of randomly chosen homologous genes can now be used to determine ($\alpha^{(1)}_1$, $\alpha^{(1)}_2$).

\begin{flushleft}
\printbibliography
\end{flushleft}

\end{document}